%% file: main.tex
\documentclass[graybox]{svmult}

\usepackage{type1cm}        
\usepackage{makeidx}         
\usepackage{graphicx}        
\usepackage{multicol}        
\usepackage[bottom]{footmisc}

\usepackage{newtxtext}       %
\usepackage{newtxmath}       

\usepackage{t1enc,dfadobe}
\usepackage{cite}
\usepackage{microtype}                 
\PassOptionsToPackage{warn}{textcomp}  
\usepackage{textcomp}                  
\usepackage{mathptmx}                  
\usepackage{times}                     
\usepackage{cite}                      
\usepackage{tabu}                      
\usepackage{booktabs}                  
\usepackage{color}
\usepackage{subfigure} 
\usepackage{comment}
\usepackage{wrapfig}

\makeindex

\graphicspath{{figs/}{figures/}{pictures/}{images/}{./}} 

\newcommand{\para}[1]        {\vspace{2mm}\noindent{\textbf{#1}}}

\usepackage{amsfonts}
\newcommand{\Xspace}{\mathbb{X}} 
\newcommand{\Rspace}{\mathbb{R}} 

\newcommand{\ALMATDA}{\textsf{ALMA-TDA\,}}

\newcommand{\imgExt}[1]{#1.pdf}

\begin{document}

\title*{Using Contour Trees in the Analysis and Visualization of Radio Astronomy Data Cubes}
\titlerunning{Contour Tree Analysis for Astronomy Data Cubes}

\author{Paul Rosen, Anil Seth, Elisabeth Mills, Adam Ginsburg, Julia Kamenetzky, Jeff Kern, and Chris R.\ Johnson, Bei Wang}
\authorrunning{Rosen et al.}
\institute{Paul Rosen \at University of South Florida, \email{prosen@usf.edu}
\and Anil Seth \at University of Utah, \email{aseth@astro.utah.edu}
\and Elisabeth Mills \at Brandeis University, \email{elisabeth.mills@sjsu.edu}
\and Adam Ginsburg \at National Radio Astronomy Observatory, \email{aginsbur@aoc.nrao.edu}
\and Julia Kamenetzky \at Westminster College, \email{jkamenetzky@westminstercollege.edu}
\and Jeff Kern \at National Radio Astronomy Observatory, \email{jkern@nrao.edu}
\and Chris R. Johnson \at University of Utah, \email{crj@sci.utah.edu} 
\and Bei Wang \at University of Utah, \email{beiwang@sci.utah.edu} 
}

\maketitle


\abstract{\input{sec-ALMA-abstract}}

\input{sec-ALMA-introduction.tex}

\input{sec-ALMA-scicase}

\input{sec-ALMA-contourtree.tex}

\input{sec-ALMA-designstudy.tex}

\input{sec-ALMA-visualization.tex}

\input{sec-ALMA-experiments.tex}

\input{sec-ALMA-discussion.tex}

\begin{acknowledgement}
This work was funded in part by a NRAO-NSF ALMA Development Grant titled  \emph{Feature Extraction \& Visualization of ALMA Data Cubes through Topological Data Analysis}. 
\end{acknowledgement}

\bibliographystyle{spmpsci}
\bibliography{main}

\end{document}

%% file: sec-ALMA-abstract.tex
The current generation of radio and millimeter telescopes, particularly the Atacama Large Millimeter Array (ALMA), offers enormous advances in observing capabilities. 
While these advances represent an unprecedented opportunity to facilitate scientific understanding, the increased complexity in the spatial and spectral structure of these ALMA data cubes lead to challenges in their interpretation.  

In this paper, we perform a feasibility study for applying topological data analysis and visualization techniques never before tested by the ALMA community.  
Through techniques based on contour trees, we seek to improve upon existing analysis and visualization workflows of ALMA data cubes, in terms of accuracy and speed in feature extraction.  
We review our application development process in building effective analysis and visualization capabilities for the astrophysicists. We also summarize effective design practices by identifying domain-specific needs of simplicity, integrability,  and reproducibility, in order to best target and service the large astrophysics community.   

%% file: sec-ALMA-introduction.tex
\section{Introduction}
\label{sec:introduction}

\begin{wrapfigure}[9]{r}{0.4\linewidth}
	\centering
	\vspace{-15pt}
	\includegraphics[width=1\linewidth]{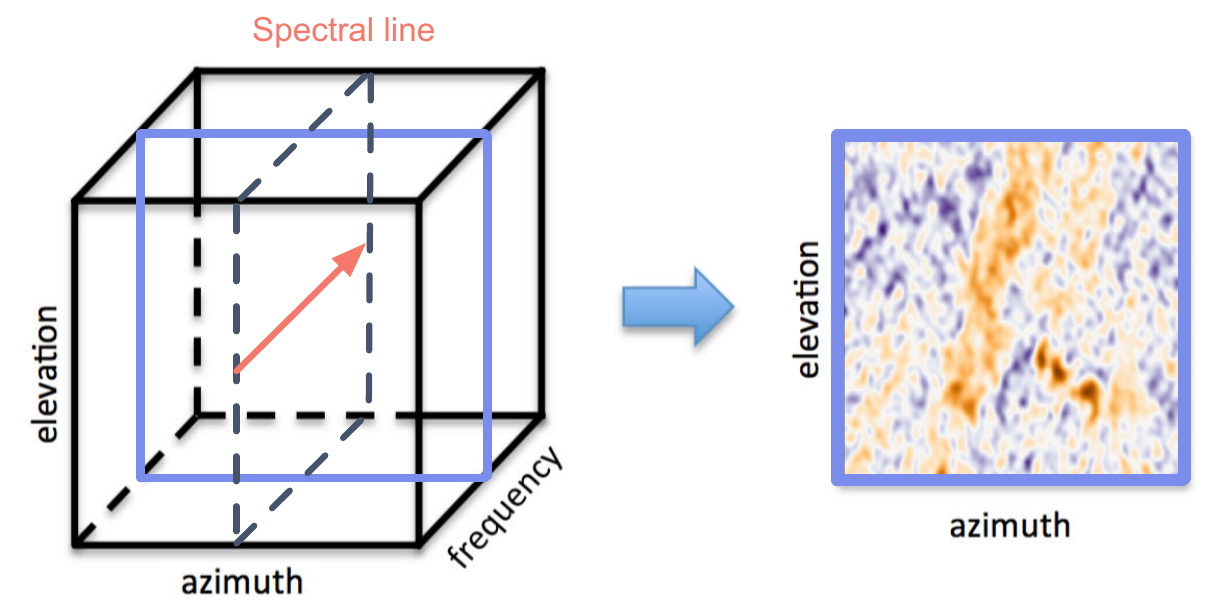}
	\caption{An illustration of the ALMA data cube and a spectral line.}
	\label{fig:cube}
\end{wrapfigure}

Radio astronomy is currently undergoing a revolution driven by new high spatial and spectral resolution observing capabilities. The current generation of radio and millimeter telescopes, particularly the Atacama Large Millimeter Array (ALMA), offers enormous advances in capabilities, including significantly increased sensitivity, resolution, and spectral bandwidth. 
While these advances represent an unprecedented opportunity to facilitate scientific understanding, they also pose a significant challenge. 
In some cases, the higher sensitivity and resolution they provide yield new detections of sources with well-ordered structure that is easy to interpret using current tools (e.g.,~\cite{PartnershipBroganPerez2015}). 
However, these advances often lead to the detection of structure with increased spatial and spectral complexity, e.g., new molecules in the chemically-rich massive star forming region Sgr B2, outflows in the nuclear region of the nearby galaxy NGC 253, and rich kinematic structure in the giant molecular cloud ``The Brick''~\cite{BellocheGarrodMuller2014, BolattoWarrenLeroy2013, RathborneLongmoreJackson2015}.  
Visualization is a natural tool to study such data, which are typically modeled as 3D cubes, commonly refereed to as \emph{ALMA data cubes}, with two spatial dimensions and one spectral dimension (see Fig.~\ref{fig:cube}).
While visualizing volumes is not new to scientific visualization, ALMA data cubes present unique challenges. 
First of all, an ALMA data cube represents the complex interactions of radio signals produced by the bulk mixing and motion of various molecules deep in space. These data tend to have
high spectral resolution but low spatial resolution. 
Yet these complex behaviors need precise examination. Second, the data have an extraordinarily low signal to noise ratio. 
This makes direct visualization impractical as the signal is difficult to extract.
Third, the noise is spectrally varying and incoherent, therefore difficult to model and remove using conventional approaches. 
Due to these unique challenges, visualization alone is insufficient for analysis and exploration.

\begin{figure}[!t]
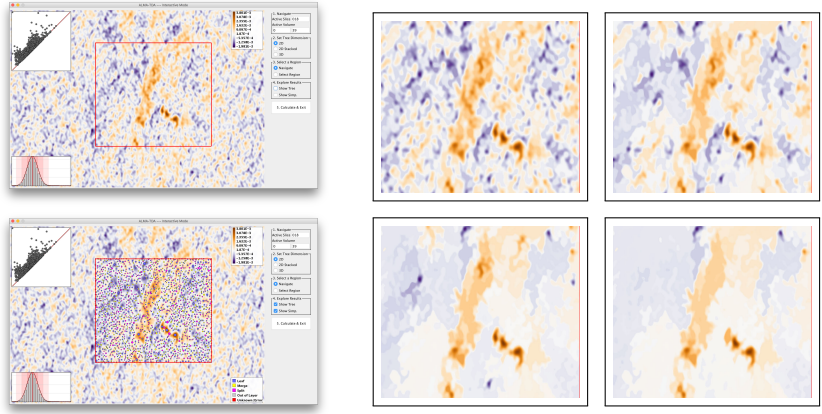

	\centering
    \begin{minipage}{0.39\linewidth}
		\centering
		\includegraphics[width=0.98\linewidth]{\imgExt{anil/vary_simp/data}}
		\includegraphics[width=0.98\linewidth]{\imgExt{anil/vary_simp/tree}}
	\end{minipage} \hspace{10pt}
    \begin{minipage}{0.51\linewidth}
		\fbox{\includegraphics[trim = 400pt 250pt 480pt 205pt, clip, width=0.44\linewidth]{\imgExt{anil/vary_simp/simp_5_0E-4}}} \hspace{2pt}
		\fbox{\includegraphics[trim = 400pt 250pt 480pt 205pt, clip, width=0.44\linewidth]{\imgExt{anil/vary_simp/simp_0_001}}}
        
        \vspace{5pt}
		\fbox{\includegraphics[trim = 400pt 250pt 480pt 205pt, clip, width=0.44\linewidth]{\imgExt{anil/vary_simp/simp_0_0015}}} \hspace{2pt}
		\fbox{\includegraphics[trim = 400pt 250pt 480pt 205pt, clip, width=0.44\linewidth]{\imgExt{anil/vary_simp/simp_0_0020}}}
	\end{minipage}
	\caption{An example of \ALMATDA. The results of varying the simplification level on slice \#18 of the Ghost of Mirach data set using a 2D contour tree. On the top left, a visualization of the original data is shown. On the bottom left, the contour tree computed on the region is shown with circles at critical point locations. Finally, on the right, the results of simplifying the data with simplification levels at $0.0005$, $0.001$, $0.0015$, $0.002$, from left to right, top to bottom, respectively.}
	\label{fig.anil.varysimp}
\end{figure}

In this paper, we review our application development process in building effective analysis and visualization capabilities for ALMA data cubes. 
Our publicly available tool is called \ALMATDA (\url{https://github.com/SCIInstitute/ALMA-TDA}). 
\ALMATDA uses contour trees to extract and simplify the complex signals from noisy radio astronomy data. 
An example of our tool is shown in Fig.~\ref{fig.anil.varysimp}.
We additionally summarize effective design practices targeting and servicing the large astrophysics community, in particular, we need to design tools with \emph{simplicity} (i.e., light-weight), \emph{integrability} (i.e., integrable within existing tool chains) and \emph{reproducibility} (i.e., fully recorded analysis history via command-lines). 
We hope such learned design practices will provide guidelines toward future development of tools and techniques that would benefit astrophysicists' scientific goals.

%% file: sec-ALMA-scicase.tex
\section{Science Case}

The new complexity involving ALMA data cubes brought about by increased sensitivity, spatial and spectral resolution, and spectral bandwidth has become a significant bottleneck in science, as it not only challenges astrophysics  analysis tools but also the users' ability to understand their data.
For example, increased complexity in the spatial and velocity
structure of spectral line emission makes even a
single spectral line hard to interpret. When a cube contains the
superposition of multiple spatial and kinematic structures, such as
outflows and rotation and infall, each with their own relationship
between the observed velocity and their actual position along the line
of sight, traditional analysis and exploration tools do not perform well. 
Users (the astrophysicists) struggle to follow kinematic trends across multiple structures by
examining movies or channel maps of the data. However, moment map
analysis (e.g., integrated fluxes, mean velocities and mean
line-widths), the most commonly used analysis tool for compressing this
3D information into a more easily parsed 2D form, no longer has a
straightforward interpretation in the presence of such complex
structure, in which mean velocities may be velocities at which no
emission is actually present, and mean line widths may represent the
distance between two velocity components, rather than the width of a
single component.

Whether scientists can navigate and correctly interpret this new
complexity will determine their success in addressing a number of
important scientific questions.  Among the topics driven by the
detection of more complex structures are ISM
turbulence~\cite{De-BreuckWilliamsSwinbank2014, RathborneLongmoreJackson2014}, the star
formation process~\cite{LiuGalvan-MadridJimenez-Serra2015},
filaments~\cite{PerettoFullerDuarte-Cabral2013}, molecular cloud structure and
kinematics~\cite{RathborneLongmoreJackson2015}, and the
kinematics of nearby galaxies~\cite{MeierWalterBolatto2015, LeroyBolattoOstriker2015,
JohnsonLeroyIndebetouw2015} and high redshift galaxies~\cite{PartnershipVlahakisHunter2015, WangWaggCarilli2013}.

An even greater challenge arises from our ability to detect an
increased number of spectral lines in more and more sources.  There
simply are no tools capable of simultaneously visualizing, comparing,
and analyzing the dozens to hundreds of data cubes for all of the
detected spectral lines in a given source. 
Standard methods that visualize the data as moment and channel maps, animate cubes as a videos or 3D models, cannot scale up to the case involving large numbers of lines, even in non-complex, well-ordered cases, 
such as rotating disks, or expanding stellar shells. 
Users become overwhelmed by, for example, comparing these typical diagnostics for
two lines, side by side or one at a time. In the richest sources with
thousands of lines, such comparisons will simply be impossible---it
becomes necessary to resort to methods that entirely throw away either
the spectral information of moment maps or the spatial
information that requires model fitting of complex spectra (such as Principle Component Analysis). 
As a result, both exploration and analysis of the astronomy data becomes not only time consuming, but potentially incomplete.

As we move into the future and the telescopes reach their full
potential, complex spatial and velocity structures will no longer be a
problem that typically occurs in a separate subset of sources than
those exhibiting rich spectra behaviors---the two problems will coexist,
compounding the highlighted issues. The visualization and analysis
challenges currently facing radio astronomy will then only grow more
pressing as the data volumes increase and the instruments grow more sensitive.

\para{Existing Tools.}
A critical aspect to the study of ALMA data cubes is the detection, extraction and characterization of objects such as stars, galaxies, and blackholes. 
\emph{Source finding} in radio astronomy is the process of detecting and characterizing objects in radio images (in the forms of data cubes), and returning a survey catalogue of the extracted objects and their properties~\cite{HancockMurphyGaensler2012, WesterlundHarrisWestmeier2012}. 
A common practice is to use a computer program (i.e., a source finder) to search the data cubes, followed by manual inspection to confirm the sources of electromagnetic radiation~\cite{WesterlundHarrisWestmeier2012}. 
An ideal source finder aims to determine the location and properties of these astronomical objects in a complete and reliable fashion~\cite{HancockMurphyGaensler2012}; while manual inspection is often time-consuming and expensive.

Several existing tools have been used in the ALMA community in terms of source finding~\cite{Hopkins2015}, including the popular ones such as clumpfind~\cite{WilliamsGeusBlitz1994}, dendrograms~\cite{RosolowskyPinedaKauffmann2008}, cprops~\cite{RosolowskyLeroy2006}, and more recent ones such as FellWalker~\cite{Berry2015}, SCIMES~\cite{ColomboRosolowskyGinsburg2015} and NeuroScope~\cite{MerenyiTaylorIsella2016}. 
Clumpfind is designed for analyzing radio observations of molecular clouds obtained as 3D data cubes; it works by contouring the data, searching for local peaks of emission and following them down to lower intensity levels~\cite{WilliamsGeusBlitz1994}. 
The dendrograms of a data cube is an abstraction of the changing topology of the isosurfaces as a function of contour level, which captures the essential features of the hierarchical structure of the isosurfaces~\cite{RosolowskyPinedaKauffmann2008}.   
The FellWalker algorithm is a gradient-tracing watershed algorithm that segment images into regions surrounding local maxima~\cite{Berry2015}. 
FellWalker provides some ability to merging clumps, therefore simplify the underlying structures, and the merging criteria shares some similarities with persistence-based simplification. However these criteria are less mathematically rigorous compared to our approach.     
SCIMES (Spectral Clustering for Interstellar Molecular
Emission Segmentation) considers the dendrogram of emission under 
graph theory and utilizes spectral clustering to find discrete regions with similar emission properties~\cite{ColomboRosolowskyGinsburg2015}.
Finally, the most recent NeuroScope~\cite{MerenyiTaylorIsella2016} (specifically targeted for ALMA data cubes) employs a set of neural machine learning tools for the identification and visualization of spatial regions with distinct patterns of motion. 

However, the study of source finding for ALMA data cubes raises the following question: How can we help the astrophysicists to understand the \emph{de-noising} process? That is, how to best separate signals from noise, and to understand the effects of de-noising on the original data? In other words, it is important for us to \emph{quantify both signals and noise} as well as to \emph{perform simplifications} of the underline data. 
This kind of study is underdeveloped with current approaches in the ALMA community.

%% file: sec-ALMA-contourtree.tex
\section{Technical Background}
\label{sec:contourtree}

From a technical perspective, we focus on performing data analysis and designing effective visualization of ALMA data cubes by
employing the contour tree~\cite{CarrSnoeyinkAxen2003}.  
The contour tree is a mathematical object describing the evolution of the level sets 
of a 
\begin{wrapfigure}[19]{r}{0.575\linewidth}
	\centering
	\vspace{-9mm}
	\includegraphics[width=1\linewidth]{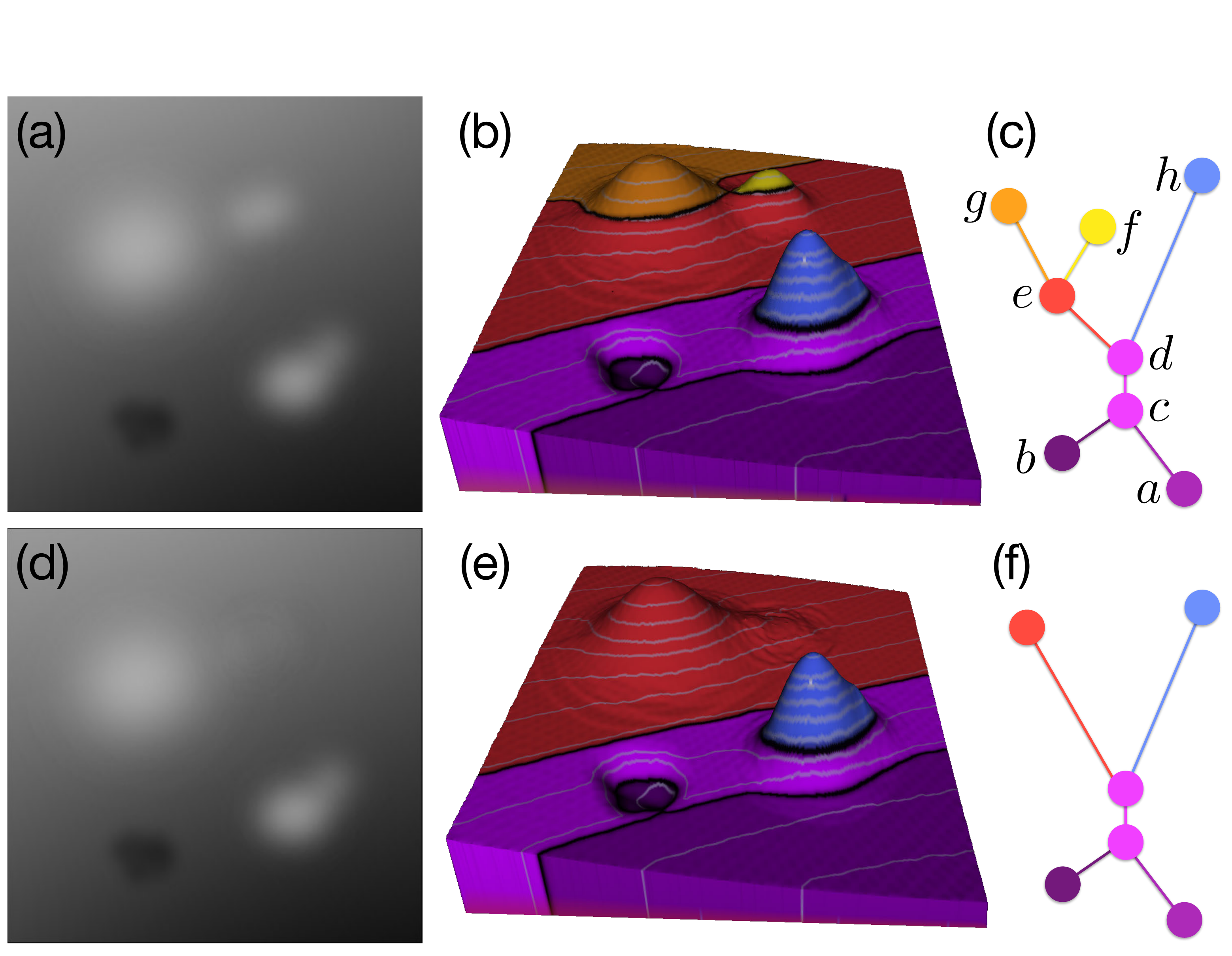}
	\caption{(a) A grayscale image of a 2D scalar function before simplification. (b) height map of the contours
corresponding to the scalar function shown in (a). (c) The contour tree structures that capture the
evolution of terrain features (i.e., relations among local minima, local maxima, and saddles).
(d)-(f): The grayscale image, height map, and the contour tree after simplifying the features.}
	\label{fig:contour-tree}
\end{wrapfigure}
scalar function defined on a simple, connected domain, such as the grayscale intensity
defined on the 
2D domain associated with a slice of a data
cube (at a fixed frequency). There are two key properties associated with a contour tree, making it a feasible tool in the study of ALMA data cubes. First, a contour tree has a graph-based representation that captures the changes within the topology of a scalar function and provides a meaningful summarization of the associated data. Second, a contour tree can be easily simplified, in a quantifiable way, to remove noise while retaining important features in data.

\begin{wrapfigure}[7]{r}{0.425\linewidth}
	\centering
	\vspace{1mm}
	{\includegraphics[trim = 20pt 20pt 20pt 20pt, clip, width=1\linewidth]{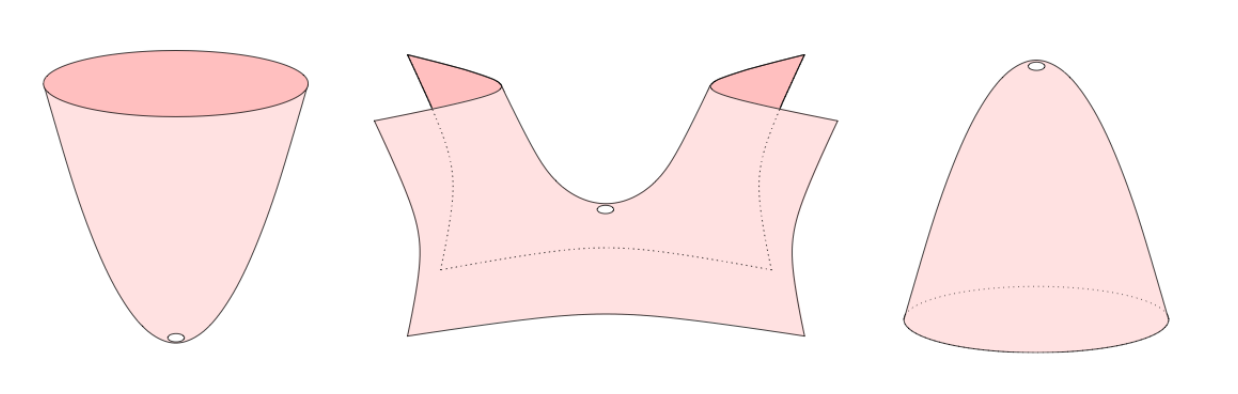}}
	\caption{Local structures of critical points. From left to right: a local minimum, a saddle point, and a local maximum.}
	\label{fig:criticalpoints}
\end{wrapfigure}

\para{Contour Trees.}
Scalar functions are ubiquitous in modeling scientific
information. Topological structures, such as contour trees, are
commonly utilized to provide compact and abstract representations for
these functions.  The contour tree of a scalar function $f: \mathbb{X} \to \mathbb{R}$ describes the connectivity of its \emph{level sets} (isosurfaces) 
$f^{-1}\left(a\right)$ (for some $a \in \mathbb{R}$), whose connected components are referred to as \emph{contours}.  Given a
scalar function defined within some Euclidean domain $\mathbb{X}$, the contour tree is constructed by collapsing the connected components of each level set to a point.  The contour tree stores information regarding
the number of components at any function value (isovalue) as well as
how these components split and merge as the function value changes.
Such an abstraction offers a global summary of the topology of the
level sets and enables the development of compact and effective
methods for modeling and visualizing scientific data.  
See Fig.~\ref{fig:contour-tree}(a)-(c) for an illustrative example. 
Vertices in the contour tree correspond to \emph{critical points} of the 2D scalar function, namely, local minima, saddle points, and local maxima, whose local structures are illustrated in Fig.~\ref{fig:criticalpoints}.

\begin{wrapfigure}[13]{r}{0.5025\linewidth}
	\centering
	\vspace{-5mm}
	\includegraphics[trim=0 0 200pt 0,clip,width=0.65\linewidth]{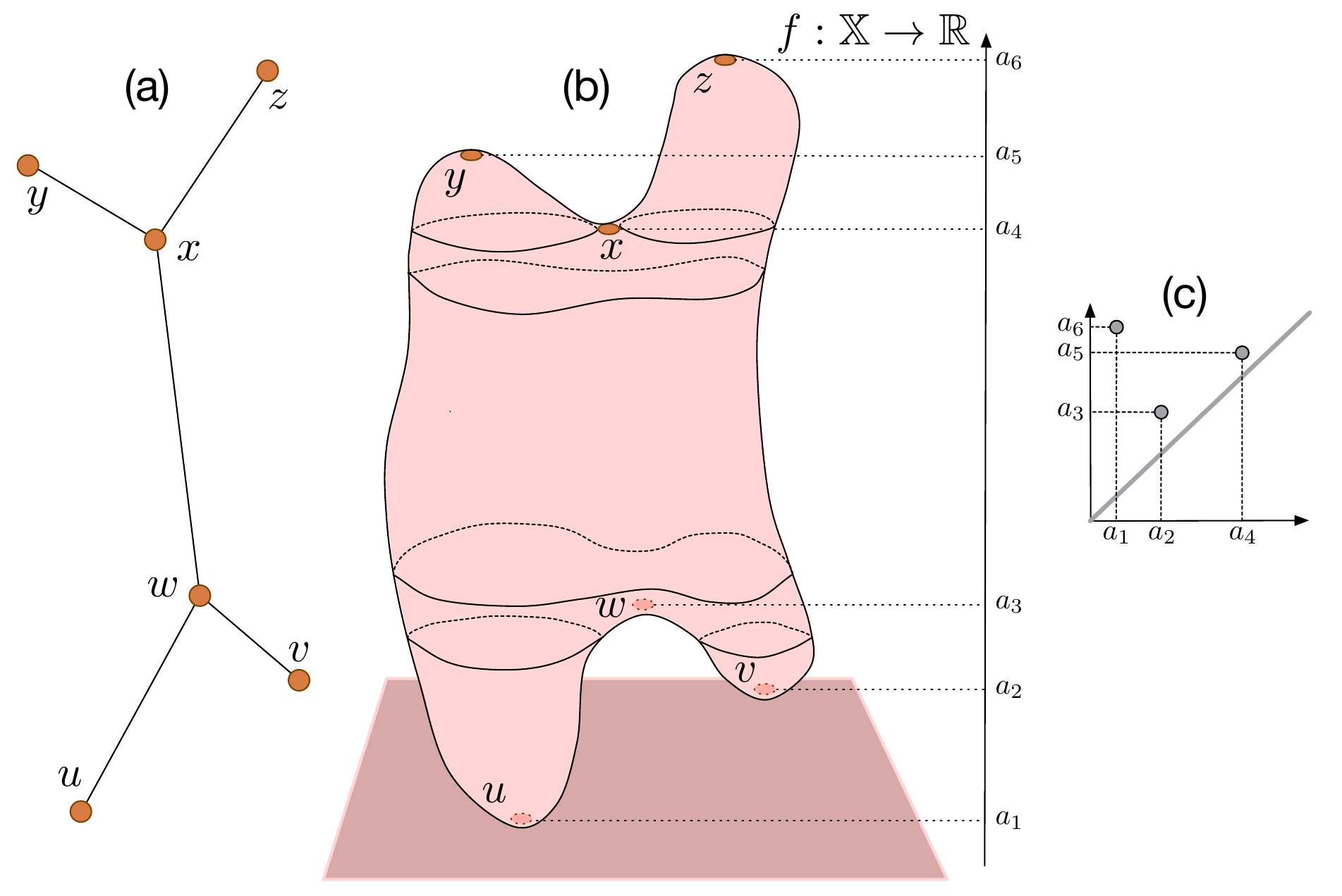}
	\includegraphics[trim=700pt 150pt 0 150pt,clip,width=0.325\linewidth]{persistence.pdf}
	\caption{Example of persistence pairing of critical points for (b) a 2D height function. The persistence pairing of (a) critical points from the contour tree gives rise to (c) a (scaled) persistence diagram.}
	\label{fig:persistence}
\end{wrapfigure}
\para{Persistence.}
To simplify a contour tree, we assign an importance measure to each edge of the tree and collapse (eliminate) edges of lower importance measures~\cite{CarrSnoeyink2003, CarrSnoeyinkPanne2004}. 
Various geometric properties, such as persistence, volume, and surface area, can be used to compute the importance measure.

We apply ideas from topological persistence~\cite{EdelsbrunnerLetscherZomorodian2002} in our feasibility study. 
We describe the idea of persistence using Fig.~\ref{fig:persistence} as an illustrative example~\cite[Page 163]{EdelsbrunnerHarer2010}. 
Given a height function $f: \Xspace \to \Rspace$ defined on a 2D domain, let $\Xspace_a$ denote the sublevel set of $f$, that is, $\Xspace_a = f^{-1}\left(-\infty, a\right]$. 
Suppose we sweep a horizontal plane in the direction of increasing height values, and keep track of the (connected) components of $\Xspace_a$ while increasing $a$. 
A component of  $\Xspace_a$ \emph{starts} at a local minimum, and \emph{ends} at a (negative) saddle point when it merges with an older component (i.e., a component that starts earlier). 
This defines a minimum-saddle persistence pair between critical vertices, and the \emph{persistence} of such a pair is the height difference between them. 
Similarly, a hole/tunnel of $\Xspace_a$ starts at a (positive) saddle point and ends at a local maxima (where it is capped off). 
This defines a saddle-maximum pair with its persistence being the height difference between its vertices.
In a nutshell, minima stars components, saddles merge components or create tunnels (complete loops), and maxima fill holes~\cite[Page 162]{EdelsbrunnerHarer2010}. 

Referring to Fig.~\ref{fig:persistence}: points $u$ and $v$ are local minima; $y$ and $z$ are local maxima; $w$ is a negative saddle point; and $x$ is a positive saddle point.  
Their corresponding height values are sorted as $a_1 < a_2 < \cdots < a_6$. 
We sweep a horizontal plane in the direction of increasing height value and keep track of the components in the sublevel set. 
The pair $(v, w)$ forms a minimum-saddle persistence pair, as a component in the sublevel set starts at point $v$ and it merges with an older component that starts at point $u$. 
The pair $(x, y)$ form a saddle-maximum pair. 
The pair $(v, w)$ has a persistence of $|a_2 - a_3|$; while the pair $(x,y)$ has a persistence $|a_5 - a_4|$. The contour tree is shown in Fig.~\ref{fig:persistence} (a).

\para{Persistence Diagram.}
The pairing of critical points also give rise to a \emph{persistence diagram}~\cite{Cohen-SteinerEdelsbrunnerHarer2007} that summarizes and visualizes topological features of a given function.  
A persistence diagram contains a multi-set of points in the plane; its $x$- and $y$-coordinates captures the start (\emph{birth}) time and the end (\emph{death}) time of a particular topological feature. The distance of the point to the diagonal captures the persistence of that feature. Points away from the diagonal have high persistence, and correspond to signals of the data; while points that are close to the diagonal have low persistence, which are typically treated as noise\footnote{Based upon sublevel set filtration, topological features typically appear as points in the upper left corner of the persistence diagram; points in the lower right corner correspond to features in superlevel set filtration and/or extended persistence~\cite{Cohen-SteinerEdelsbrunnerHarer2009}, which are not the focus of this paper.}.
 
In the example of Fig.~\ref{fig:persistence}, the critical point pairs $(x,y)$ and $(v, w)$ give rise to points $(a_4, a_5)$ and $(a_2, a_3)$ in the persistence diagram, respectively. 
This persistence diagram also contains an additional off-diagonal point $(a_1, a_6)$, which corresponds to the pairing of global minimum $u$ with the global maximum $z$ that captures the entire shape of data. This is a global feature that can not be simplified (see~\cite{Cohen-SteinerEdelsbrunnerHarer2009} for technical details). 
In our context, we only care about minimum-saddle and saddle-maximum pairs. 

\para{Contour Tree Simplification.} 
In the contour tree example of Fig.~\ref{fig:contour-tree}~(c), the pair $(b, c)$ is a minimum-saddle pair while the pairs $(e, f)$ and $(d, g)$ are saddle-maximum pairs. 
During a hierarchical simplification, the pair $(e,f)$ has the smallest persistence, therefore the edge connecting them is collapsed (simplified), as shown in Fig.~\ref{fig:contour-tree}(f); this can be achieved by a smooth deformation of the surface in Fig.~\ref{fig:contour-tree}(d). 
In this paper, we focus on the persistence-based simplification, other simplification schemes may be employed based on local geometric measures for individual contours~\cite{CarrSnoeyinkPanne2004}, for instance, surface area and contained volume; we intend to use these geometric measures in the future to perform contour tree simplification that suppressing minor topological features of the astronomy data.

\para{Scalar Field Simplification.} 
Given a contour tree simplification, we would like to compute its corresponding scalar field simplification. 
Simplifying a scalar function directly in a way that removes topological noise as determined by its persistence diagram has been investigated extensively (e.g.~\cite{EdelsbrunnerMorozovPascucci2006}). 
As pointed by Carr at al.~\cite{CarrSnoeyinkPanne2010}, contour tree simplification have well-defined effects on the underlying scalar field: collapsing a leaf corresponds to leveling off (or flattening) regions surrounding a maximum or a minimum. 
This is a desirable simplification for the domain scientists, as they are interested in reducing noise to zero flux during the de-noising process. 
Fig.~\ref{fig:contour-tree}(b) and (c) demonstrate the result of edge collapsing: collapsing the edge $(e, f)$ from the tree results in flattening  the yellow region surrounding the local maximum~$f$; 
\begin{wrapfigure}[18]{r}{0.437\linewidth}
	\centering
	\vspace{4mm}
	\includegraphics[trim=0 0 0 10pt, clip, width=1\linewidth]{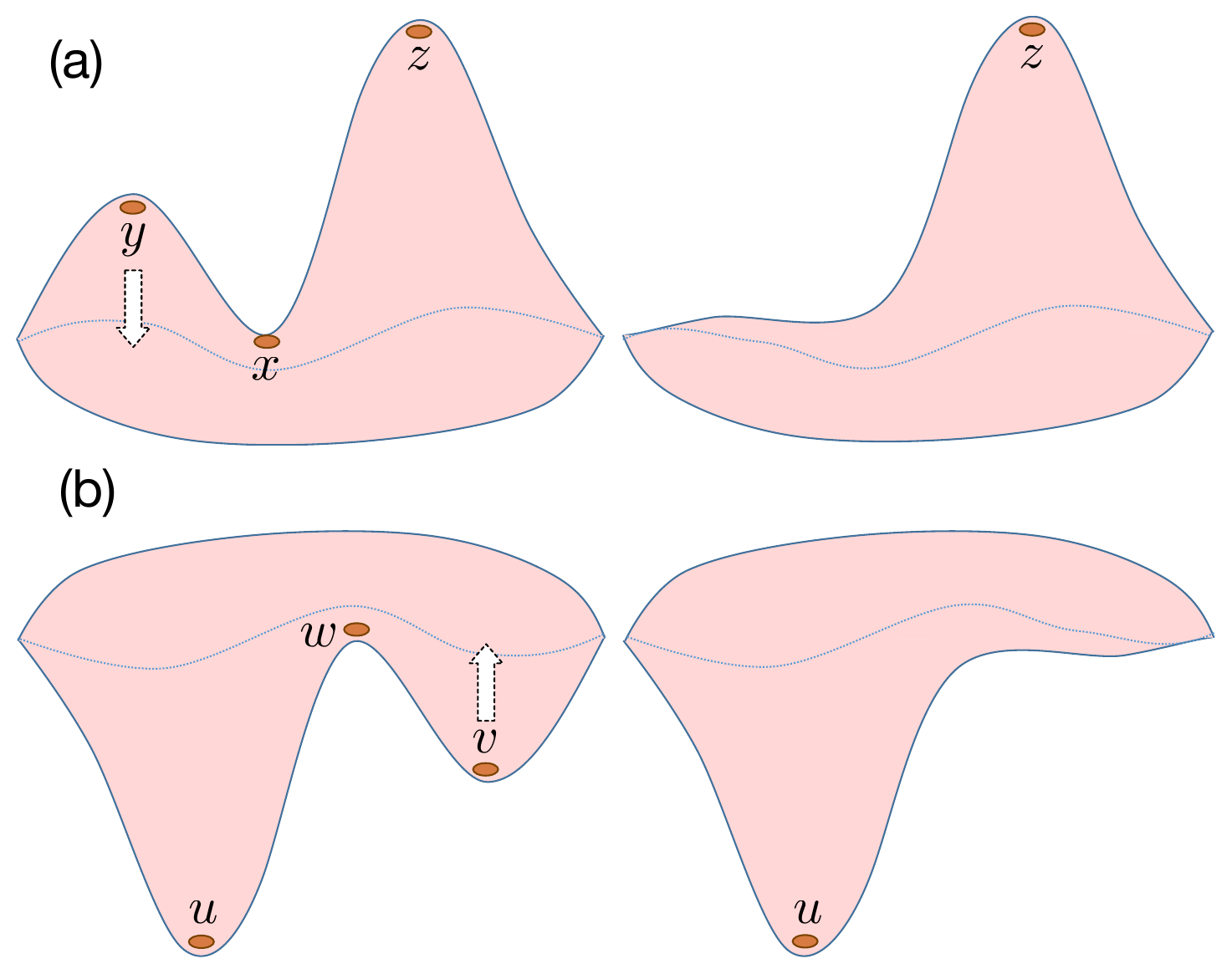}
	\caption{Simplifying a saddle-maximum pair $(x, y)$ in (a) and minimum-saddle pair $(v, w)$ in (b) for a 2D scalar field. 
(a): We reduce the height of the local maximum $y$ to the level of saddle $x$, effectively flattening the region surrounding $y$. (b): We increase the height of local minimum $v$ to the level of the saddle $w$, again flattening the region surrounding~$v$.}
	\label{fig:simplify}
\end{wrapfigure}
this is equivalent to introducing a small perturbation to the neighborhoods of saddle-maximum pair $(e, f)$ so that both critical points $e$ and $f$ are removed. 
Such a flattening process is further highlighted in Fig.~\ref{fig:simplify}.


\para{Related Work.}
The contour tree was first introduced by van Kreveld et al.~\cite{KreveldOostrumBajaj1997} to study contours on topographic
terrain maps (i.e., curves containing sampled points with the same
elevation values).  It has then be widely used for both scientific and
medical visualizations~\cite{BajajPascucciSchikore1997, PascucciCole-McLaughlinScorzelli2005, SchneiderWiebelCarr2008, SohnBajaj2006}.
Efficient algorithms for computing the contour tree~\cite{CarrSnoeyinkAxen2003, ChiangLenzLu2005,RaichelSeshadhri2014} (and its variants, merge tree~\cite{OesterlingHeineWeber2015},
and Reeb graph~\cite{PascucciScorzelliBremer2007}) in arbitrary dimensions have been developed.
Calculation of contour trees is theoretically $\mathcal{O}\left(n\log{}n\right)$. However, the actual running time is approximately $\mathcal{O}\left(n\right)$.
The latest state-of-the-art regarding contour trees have been parallel or distributed implementations~\cite{MorozovWeber2013,MorozovWeber2014, CarrSewellLo2016,CarrWeberSewell2016,MaadasamyDoraiswamyNatarajan2012,GuenetFortinJomier2016}. We use an approach described in~\cite{rosen2017ct}, which is implemented under the piecewise linear setting.

A related concept called dendrograms has been used in astronomical
applications to segment data and to quantify hierarchical structure
such as the amount of emission or turbulence at a given size scale,
for example, to study the role of self-gravity in star
formation~\cite{GoodmanRosolowskyBorkin2009}.  A dendrogram is a
tree-diagram typically used to characterize the arrangement of
clusters produced by hierarchical clustering. It tracks how
components (clusters) of the level sets merge as the function value
changes, while a contour tree captures more complete topological
changes (i.e., merge and split) of the level sets.  The
state-of-the-art Astronomical Dendrogram method~\cite{Astrodendro} has limited capabilities in automatic data denoising, feature extraction and interactive visualization.

%% file: sec-ALMA-designstudy.tex
\section{Application Development Process}
\label{sec:design-study}

We revisit our application development process in building effective analysis and visualization capabilities of ALMA data cubes by reviewing the timeline of our project. We reflect on key activities with the goal of learning from experience and summarizing effective design targeting and serving the astrophysics community, and how different members of our team interact with one another, including computer scientists (both visualization and TDA experts) and radio astronomers.  
%
%
To give an overview of our design process, we describe the critical activities as identified in~\cite{McKennaMazurAgutter2014}: understand, ideate, make, and deploy. 

The discussion of the project started in November 2014, when National Radio Astronomy Observatory (NRAO) scientist Dr.\ Jeff Kern, a coauthor of this paper, visited the Scientific Computing and Imaging (SCI) Institute and saw a talk on the topic of topological data analysis. Over the following months, follow-up conversations generated some initial excitement regarding the potentially applying topological techniques in understanding ALMA data cubes, which has never been done before.

To \emph{understand} the problem domain and target users, we identified key opportunities, that is, applying emerging techniques from topological data analysis to the study of ALMA data cubes. The main motivation stemmed from Jeff's comments that ``there simply are no tools capable of simultaneously visualizing, comparing, and analyzing the dozens to hundreds of data cubes for all of the detected spectral lines in a given source.'' We believed that introducing topological data analysis techniques to the ALMA community would potentially offer new insights regarding feature detection, as well as improve their workflow efficiency. 

The \emph{ideate} activity of the project started in May 2015, as the domain problems became better characterized and possible solutions via contour tree-based approaches appeared to have the greatest potential among the solution space.  We externalized our ideas and expected technical challenges, while at the same time, formulating a potential analysis pipeline, visual encodings, and selecting interactive capabilities within a proposed system for ALMA data cubes. 

By January 2016, we have already met with astrophysicists at NRAO facility to learn their needs and conducted an on-campus interview with astrophysicist, Dr.~Anil Seth, another collaborator of this project, who works with ALMA data cubes. 
We learned the typical pipeline in the analysis and visualization of ALMA data cubes, specifically, in Anil's case, via image editing tools or file viewers such as QFitsView~\cite{QFitsView} and SAOImage DS9~\cite{DS9}. 
We also gave short tutorials regarding our proposed techniques to obtain comments and feedback from all our interactions. 

We started our \emph{make} activity by constructing a tangible prototype, specifically encompassing visualization decisions and interaction techniques. The process coupled the ideate and make activities in the design and refinement of our system. We identified that \emph{quantification} (of signals and noise) and \emph{simplification} are two of the most important aspects for our proposed framework. We went through multiple rounds of interface mock-ups and functionality discussions. We showcased our first prototype between June and August 2016, including one-on-one discussions with Anil and Dr.\ Julia Kamenetzky on our team, and through a number of talks given to the astrophysics community, with general positive feedback. 

Over the course of the next half an year, we rolled out multiple phases of \emph{deploy} activities, in order to put the prototype in real-world setting to understand how to improve its effectiveness and performance. Our goal was to have a usable system that helps with the users' data-specific tasks. 
In January 2017, we organized a one-day workshop where we engaged in panel discussions on the current version of the prototype, gathered comments and suggestions, and discussed potential research and developmental directions moving forward. This workshop in particular helped to cement the lessons learned. 

\subsection{Designing to Serve the ALMA Community} 

Throughout the development process, we learned a few best practices for serving the ALMA community: simplicity, integrability, and reproducibility. 

In terms of \emph{simplicity}, the tool should contain sufficient but not overwhelming amount of visualization; and minimize GUI interactions. This philosophy is in sharp contract with some of the common practices of many visualization tools, where we aim to create novel, exciting and sometimes flashy visualizations. Our initial prototypes were full of many unnecessary functionalities and complex GUIs. We learned via feedbacks and user practices that a complex interface will distract or confuse the users to the point that they would not even try using the software. 
The tool should also be light-weight. That is, it should be easily installed on a desktop computer and not require extensive external dependencies or packages be installed. For this reason, we chose Java as a platform from the beginning. Though not the most efficient, Java software is highly portable. This is well-aligned with properties of commonly used processing tools in the ALMA community. 

In terms of \emph{integrability}, the tool should be integrable with existing workflows and toolchains. This means that the core functionality of the software need to be automatable. In addition to providing a GUI, we also provide a command-line interface for generating results, such that it can one day be integrated with other tools such as CASA (Common Astronomy Software Applications)~\cite{mcmullin2007casa}, astropy~\cite{robitaille2013astropy}, or SAMP (Simple Application Messaging Protocol)~\cite{taylor2015samp}.

In terms of \emph{reproducibility}, the analysis history using our tool should be recorded so that the results can be reproduced. 
This is supported in two ways. First, by enabling processing via the command-line, we can save parameters and automatically rerun the results later. Second, we minimize the amount of GUI interactions, as most of such interactions are exploratory and do not necessarily contribute to the final analysis. When the user is satisfied with their results using our visualization, the exact command required to reproduce the visualization results is output to the command-line for future reference.









%% file: sec-ALMA-visualization.tex
\section{Software Design}
\label{sec:visualization}

Our software is a visualization tool with both command-line only and interactive visualization operating modes. 

The command-line mode provides a small set of options for complete reproducibility of any computation. 
Those options are:
\begin{itemize} 
\item \textbf{Input file} -- Path to the file for processing.
\item \textbf{X, Y, \& Z range} -- The dimensions of the region to be processed. 
\item \textbf{Simplification type} -- Either 2D, for a single slice; 2D Stack, for a series of 2D slices; or 3D, for volume processing.
\item \textbf{Simplification level} -- Persistence level for feature simplification.
\item \textbf{Output file} -- Path to save results.
\end{itemize}

We also provide an interactive visualization mode to explore the capabilities of our approach and select these parameters. When starting the software, users need only add the ``interactive'' tag to the command-line, and the visualization launches. 

\begin{figure}[!b]
	\centering
	\includegraphics[width=0.48\linewidth]{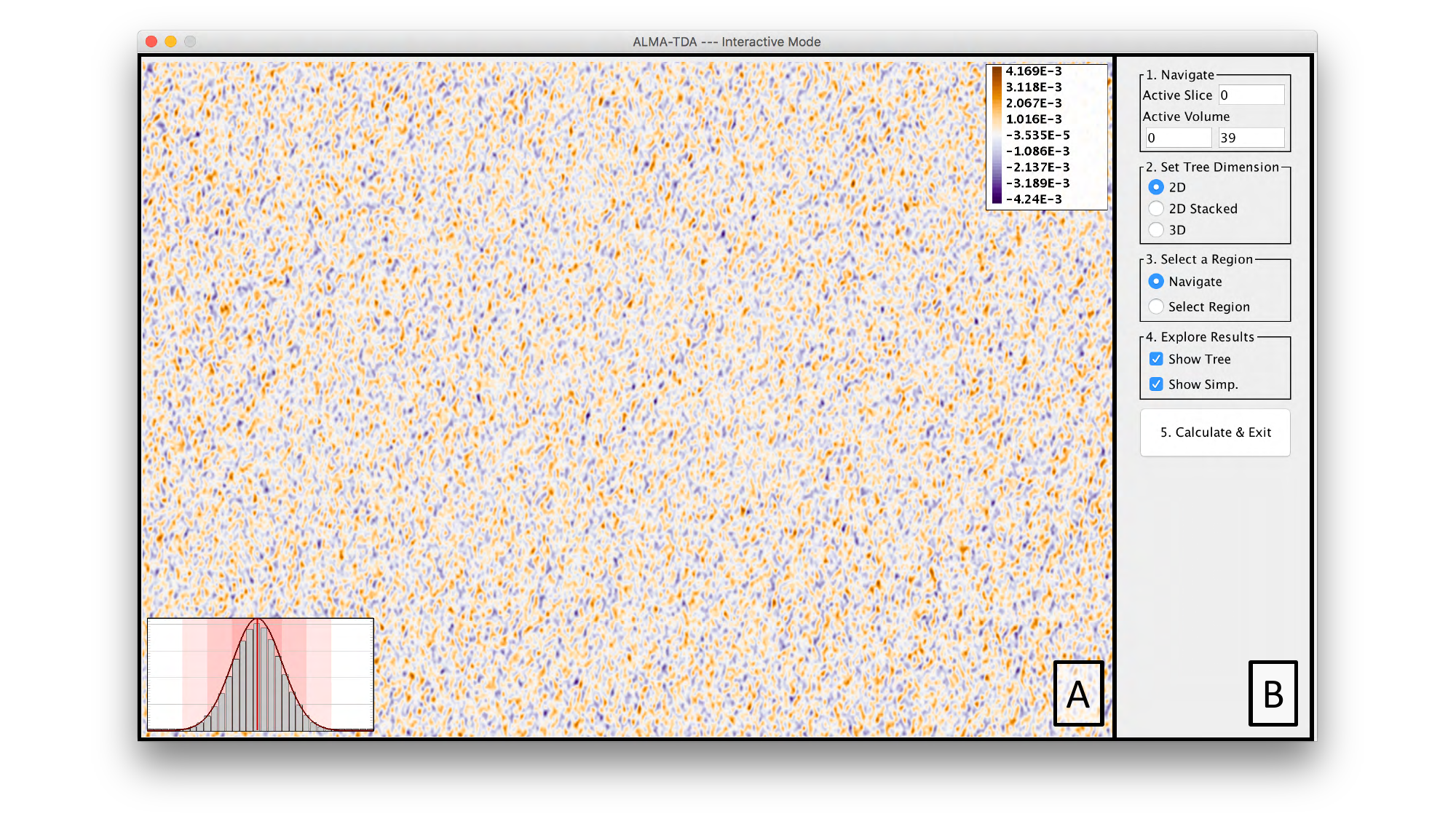}
	\includegraphics[width=0.48\linewidth]{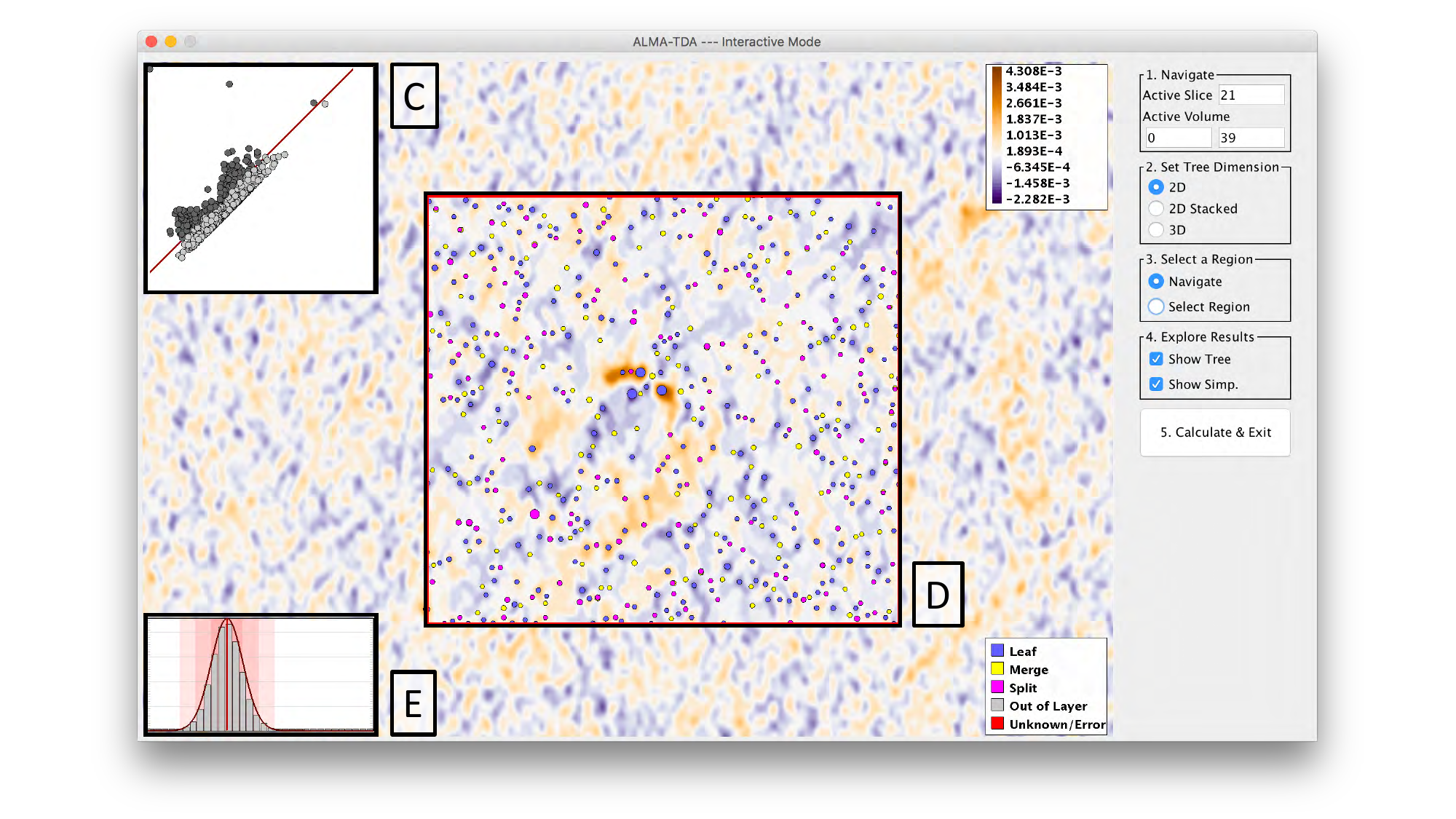}
	\vspace{-2mm}
	\caption{Upon loading the software in the interactive mode, the user is presented with a view of the data. Top: Initial view of the software. Bottom: Visualizations shown as the user selects a region of interest and the contour tree is calculated (on the back end).}
	\label{fig.software}
\end{figure}

The visualization initially opens to the interface seen in Fig.~\ref{fig.software} (left). The interface is designed to include only minimal required capabilities. The main window, \textbf{A}, shows visualizations related to the loaded data cube. The GUI component, \textbf{B}, provides controls to set options for processing the data. The controls are placed in groups, numbered for steps 1-5. The GUI component is designed with both simplicity and functionality in mind, to offer the users most intuitive, and yet fully-functional analysis capabilities. 

\subsection{Visual Elements}

The visualization is composed of five main visual elements.

\para{Scalar Field View (Fig.~\ref{fig.software} A).} 
Being a sampling of radio waves, the 2D scalar field (a slice of ALMA data cube along the frequency axis) has both positive and negative amplitudes. It is therefore displayed using a divergent orange/purple colormap. By default, the first slice is selected and viewed by centering on the middle of the domain. The user can translate and zoom with the mouse. Different slices can be selected by changing the values in the controls in Fig.~\ref{fig.software} B.

\para{Persistence Diagram (Fig.~\ref{fig.software} C).} 
Once the contour tree is calculated, the data are displayed using a persistence diagram. 
Being that the distance from the diagonal is an analog to persistence, we use this visualization for interactively selecting the level of simplification by dragging the red simplification bar. Features below the bar are grayed out indicating that those features will be simplified. Once released, a simplified contour tree (on the back-end) and a simplified scalar field (for the front-end) are calculated. 

\para{Contour Tree (Fig.~\ref{fig.software} D).} 
Displaying the tree structure of the contour tree is not particularly meaningful, as it is both large and an abstract view of the data. However, seeing critical points and their persistence in the context of the data is valuable. The critical points are placed over the scalar field view at their respective spatial location. Their size is set based upon their persistence (higher persistence, larger point). Finally, their color is set by their type: local extrema (leaf of the tree) -- blue, negative saddle points (merge) -- yellow, and positive saddle points (split) -- magenta. For 3D analysis, contour tree nodes off layer are colored gray. 
This view of the contour tree can be enabled or disabled on demand using the controls in Fig.~\ref{fig.software} B.

\para{Simplified Scalar Field (Fig.~\ref{fig.software} D).}
Since users are in large part interested in the feature extraction power of this approach, we show the result of scalar field simplification in context. As the user adjusts the level of persistent simplification, the scalar field is simplified and overlaid with the original visualization. This view can be enabled or disabled on demand using the controls in Fig.~\ref{fig.software} B.

\para{Histogram (Fig.~\ref{fig.software} E).} 
A histogram is produced, indicating the distribution of (intensity) values of data cubes within the current view. In addition to showing histogram bins, this view shows the mean as a solid red line and $\pm3$ standard deviations as consecutively lighter red bars. This histogram is adapted as the user navigates their view or when the simplification level of the scalar field is adjusted. This view is important, as domain experts are interested in quantifying the total flux gained or lost during simplification. This is most observable by shifts in the mean.

\subsection{Interaction Process}
Though the use of our software requires some explanation, we strive to make it as simple to use as possible. Part of this effort is providing a simple and intuitive five step approach to the users.

\para{Step 1: Navigation.}
The users are first asked to navigate the view to the general region of interest. This includes translation and zooming, but it also includes selecting the slice or volume of interest.

\para{Step 2: Tree Dimension.}
The dimension of the contour tree calculation must be selected next. The options include 2D, for a single slice; 2D stack, for 2D computation on a series of slices; and 3D for computation on a volume. These options will be discussed further in the case studies.

\para{Step 3: Region Selection.}
Next the specific region of interest is selected with the mouse. As soon as the mouse is released, computation begins. If the region is large, the user is prompted with the option to cancel, due to computation time. We are actively investigating scalable contour tree computations to support larger data cubes with on-the-fly visualization. 

\para{Step 4: Exploration.}
Once the computation is completed, the user is invited to explore the domain. This includes navigation (translation, zooming, and changing slices) and adjustment of the simplification level. As simplification levels are adjusted, the user can observe changes in the scalar field, compare those changes to the original field, and look for changes in flux in the histogram.

\para{Step 5: Compute and Exit.}
Steps 1-4 may be repeated as many times as necessary, until the user is satisfied. Once done, the user clicks ``Compute and Exit''. This will trigger a processing of the data cube and saving of output. Finally, the precise command required to reproduce the results will be printed on the command-line.

%% file: sec-ALMA-experiments.tex
\section{Case Studies}
\label{sec:experiments}

We show the capabilities of our prototype with two case studies involving specific ALMA data cubes used by coauthors.

\begin{figure}[!b]
	\centering
	\fbox{\includegraphics[trim = 370pt 225pt 465pt 205pt, clip, width=0.225\linewidth]{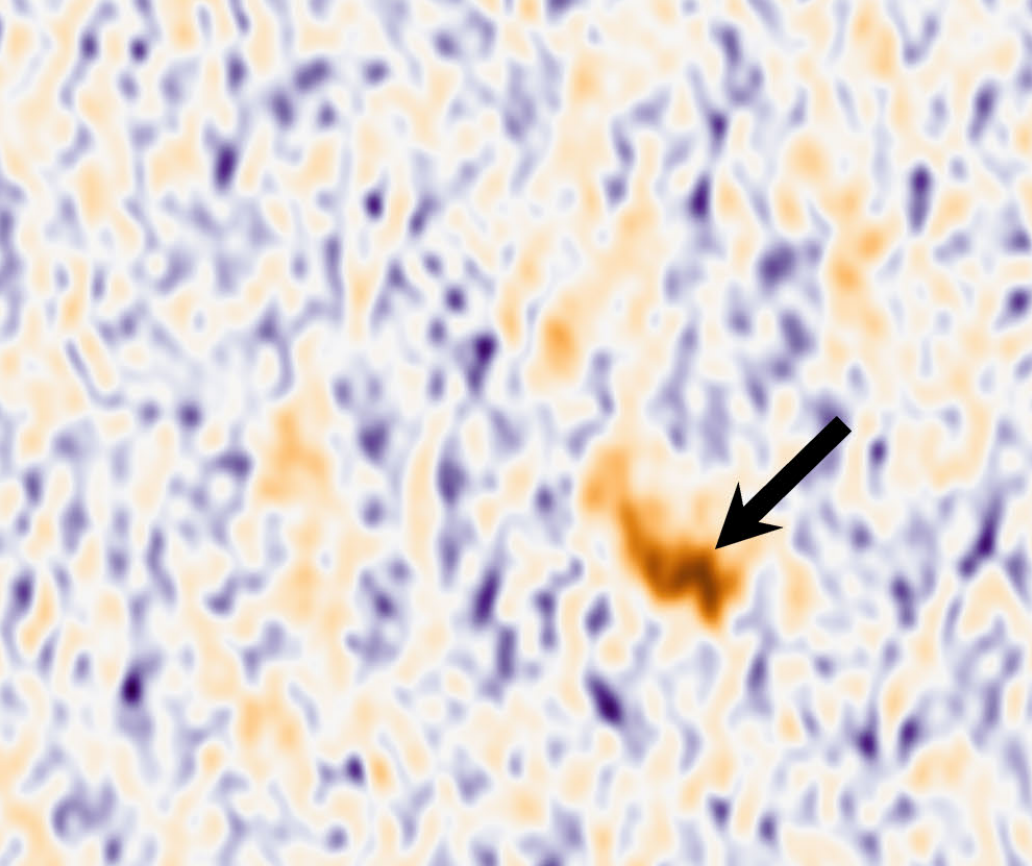}} \hfill
	\fbox{\includegraphics[trim = 370pt 225pt 465pt 205pt, clip, width=0.225\linewidth]{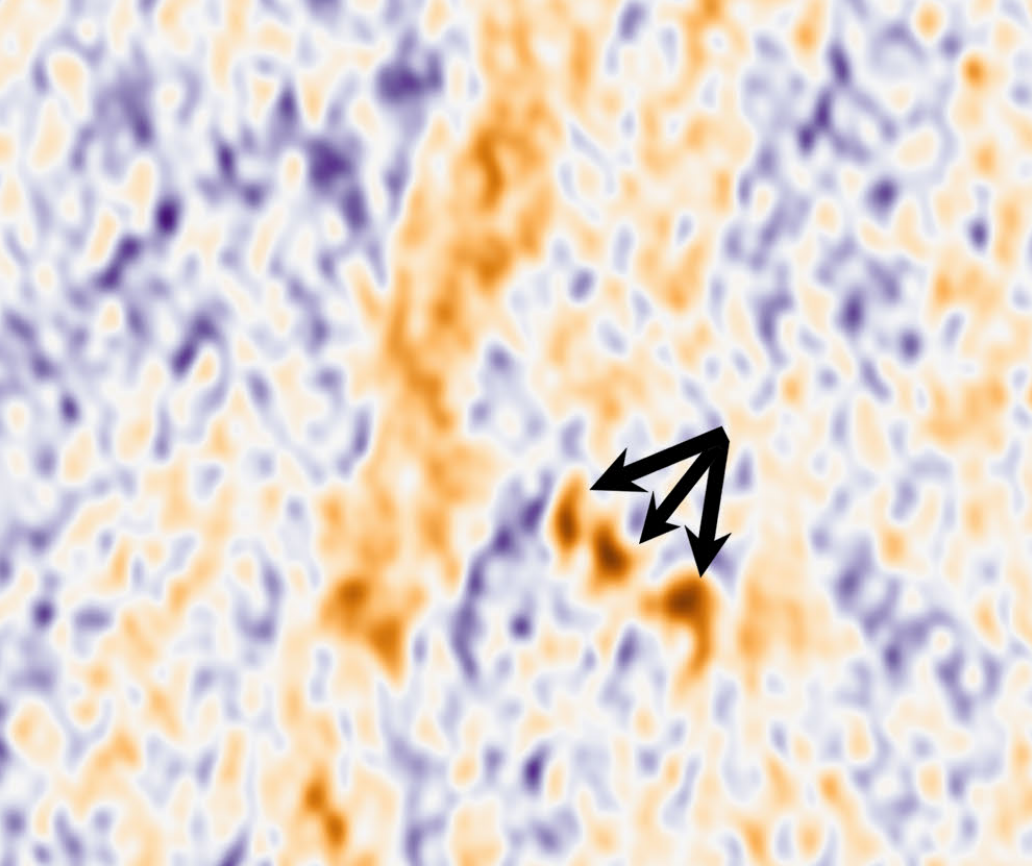}}
	\fbox{\includegraphics[trim = 370pt 225pt 465pt 205pt, clip, width=0.225\linewidth]{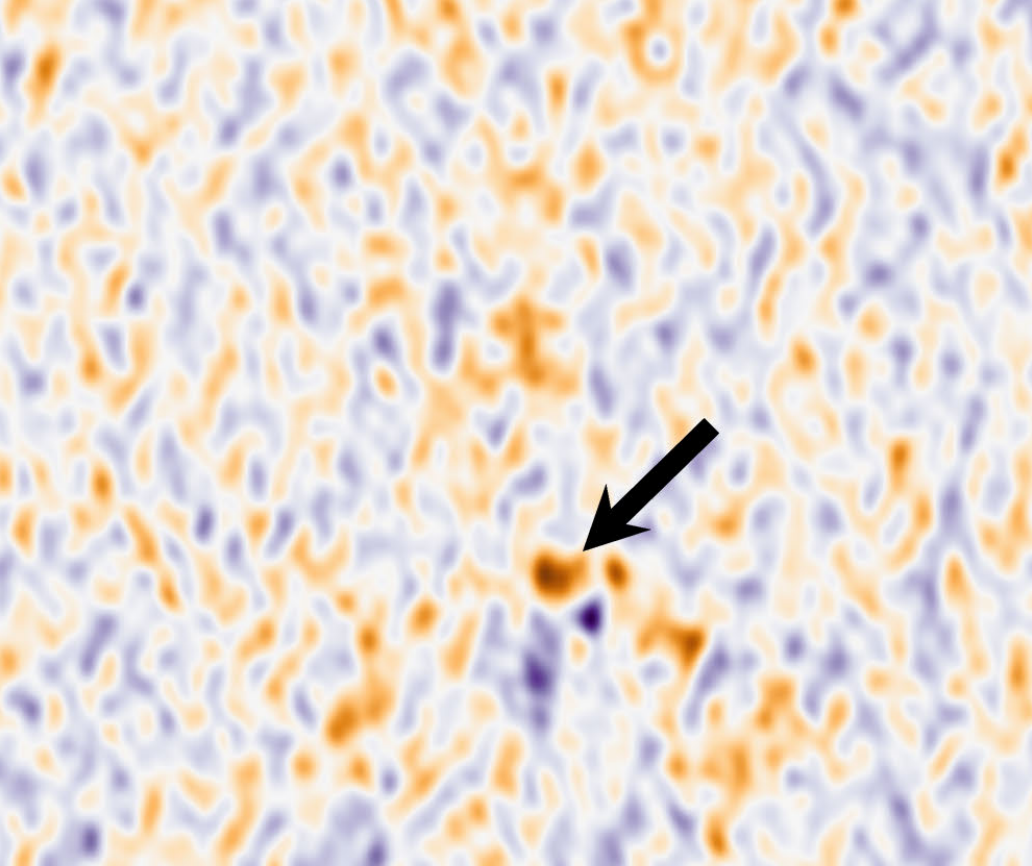}} \hfill
	\fbox{\includegraphics[trim = 370pt 225pt 465pt 205pt, clip, width=0.225\linewidth]{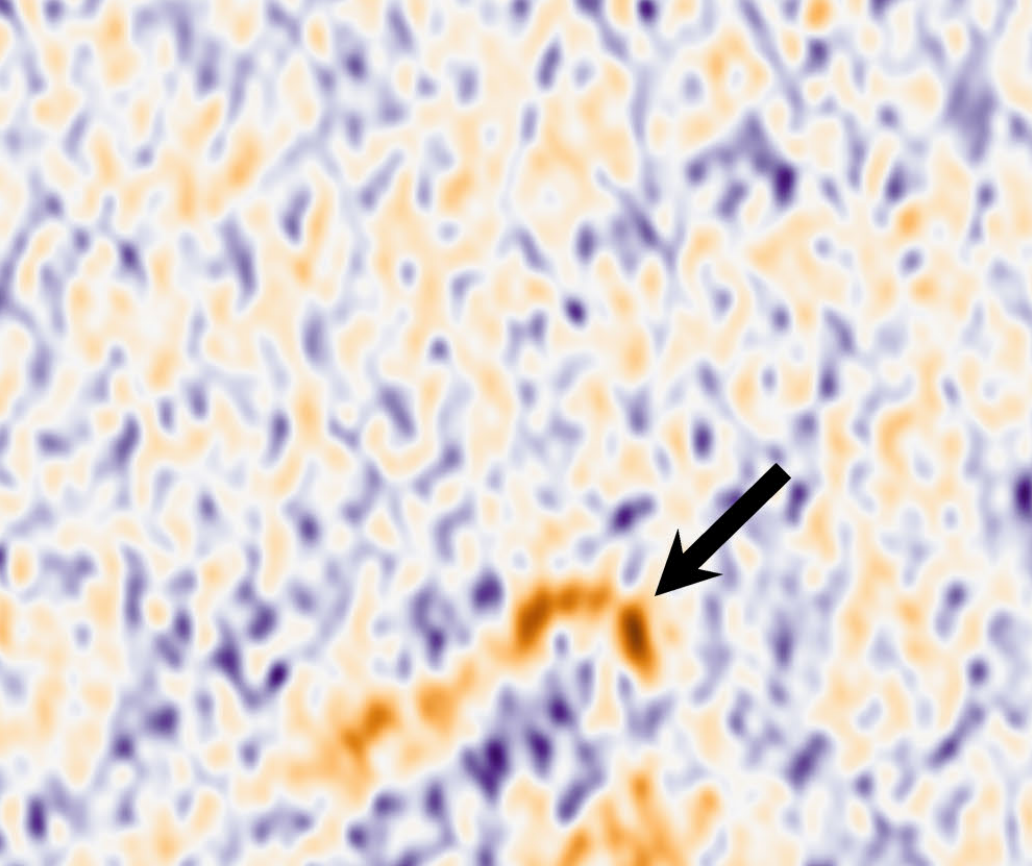}}
	\caption{Four slices, \#16, 18, 20, \& 22 from the Ghost of Mirach data set. The bright red spots (indicated by the arrows) in these images are the signal of interest.}
	\label{fig.anil.original}
\end{figure}


\subsection{Ghost of Mirach Galaxy Data Set}

NGC 404 (also known as Mirach's Ghost) is data of a molecular gas emission at the center of the nearby, low mass galaxy. The data was taken using ALMA on Oct.~31, 2015.  A data cube is created using the default ALMA pipeline and involves Fourier transformation of the interferometric data at each frequency. The data cube is approximate 4.5GB with resolution of 5400x5400 in the spatial domain and 30 in the spectral domain (i.e., 30 slices). However, the feature of interest is around 200x200 in size and covers around 10 slices. Scientists often sample cubes much larger than their feature of interest to reduce some structured errors, vignetting for example.

\begin{figure*}[!b]
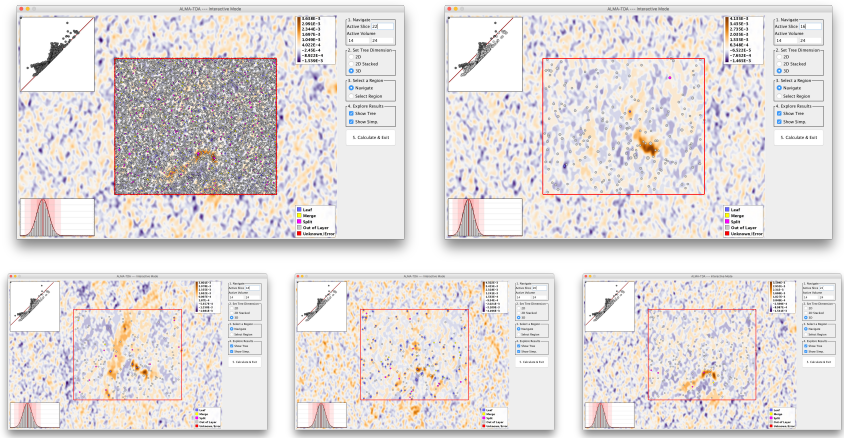

	\centering
	\includegraphics[width=0.48\linewidth]{\imgExt{anil/3dsimp_0.00128/tree_22}}
	\includegraphics[width=0.48\linewidth]{\imgExt{anil/3dsimp_0.00128/simp_16}}
	\includegraphics[width=0.32\linewidth]{\imgExt{anil/3dsimp_0.00128/simp_18}}
	\includegraphics[width=0.32\linewidth]{\imgExt{anil/3dsimp_0.00128/simp_20}}
	\includegraphics[width=0.32\linewidth]{\imgExt{anil/3dsimp_0.00128/simp_22}}
	\caption{Result of simplifying using the 3D contour tree on the Ghost of Mirach data set is worse than expect due to topological pants (tubes connecting through slices). Top left: Visualization of the 3D contour tree on slice 22. Top right: Simplification of slice 16. Bottom: Simplification of slices 18, 20, \& 22, respectively. The persistent simplification level is $0.00128$.}
	\label{fig.anil.3dvolume}
\end{figure*}

\begin{figure}[!b]
    \centering
    \includegraphics[trim=0 0 75pt 0, clip,width=0.31\linewidth]{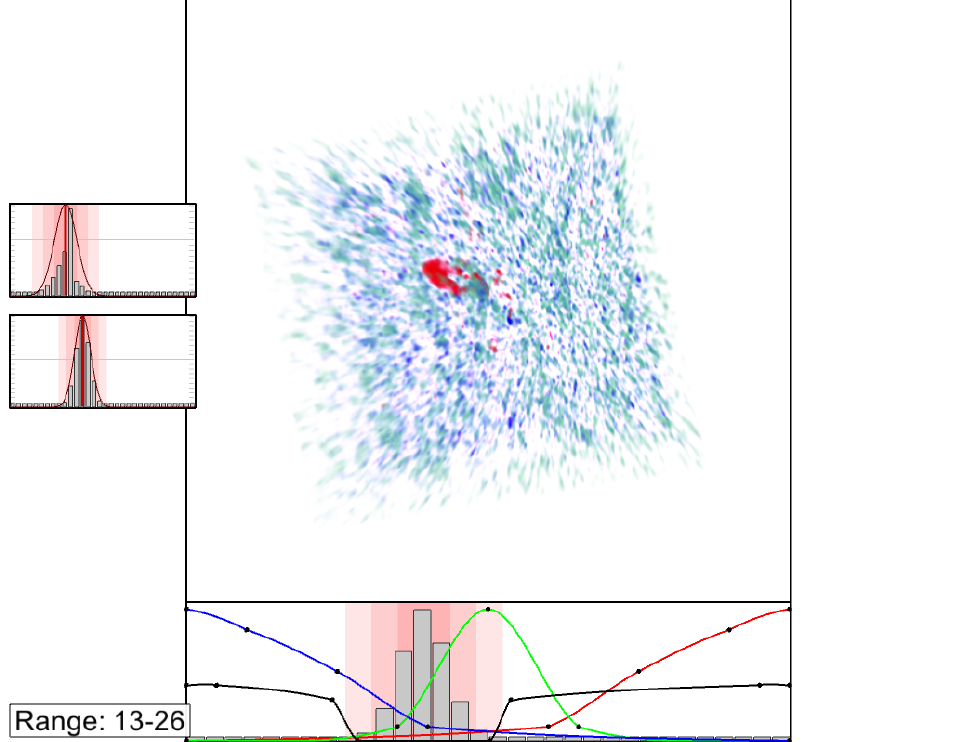}\hfill
    \includegraphics[trim=0 0 75pt 0, clip,width=0.31\linewidth]{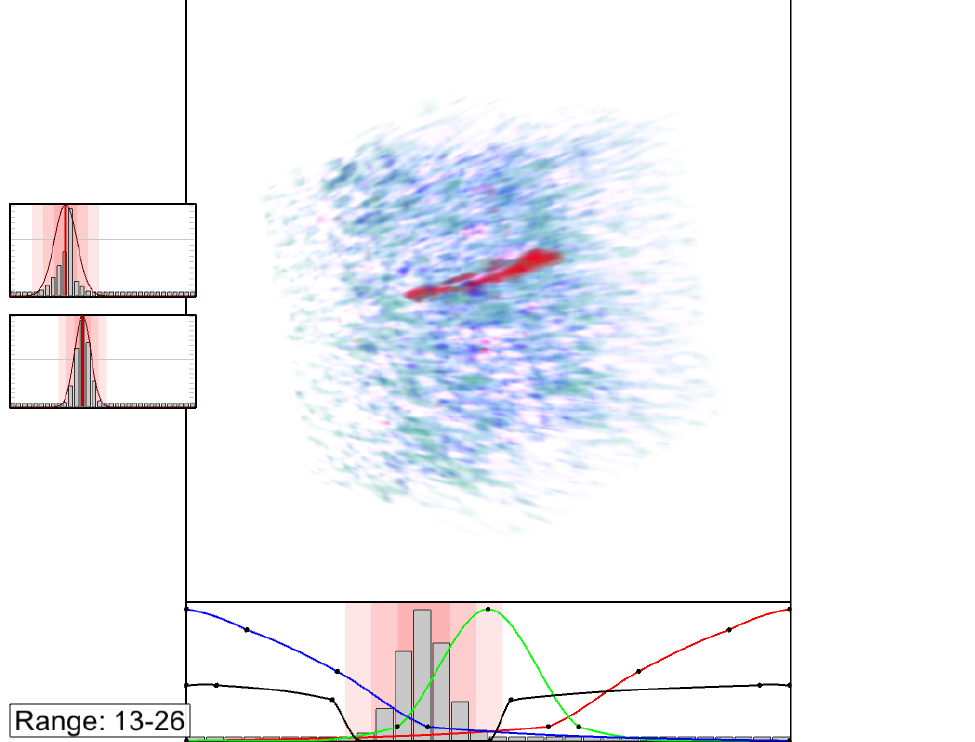}\hfill
    \includegraphics[trim=0 0 75pt 0, clip,width=0.31\linewidth]{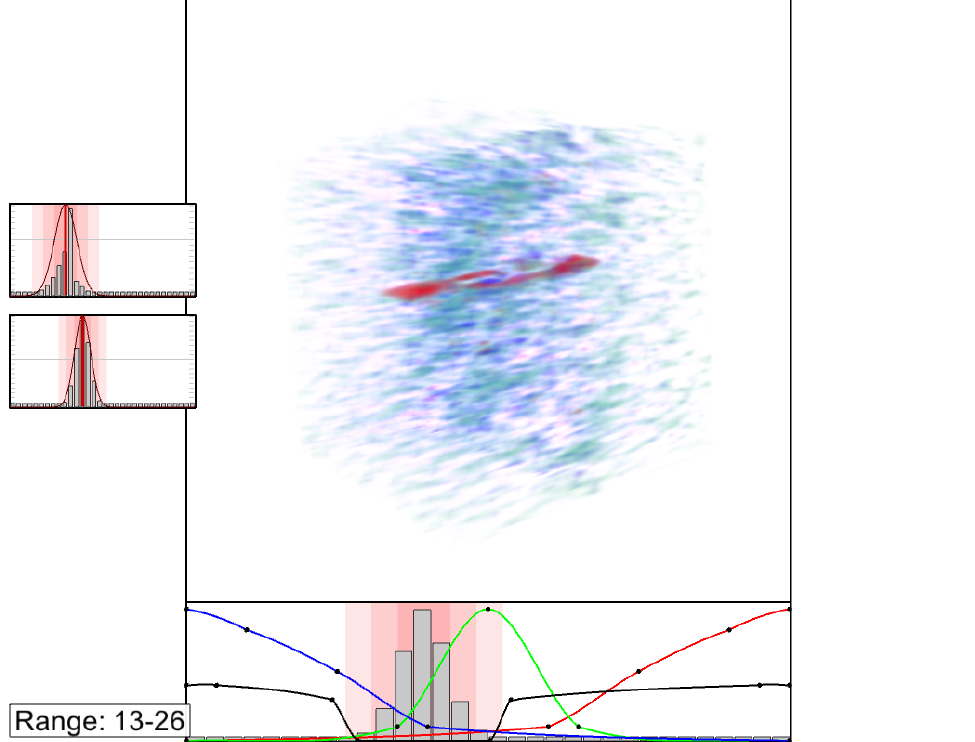}
    
    \vspace{3pt}
    \includegraphics[trim=0 0 75pt 0, clip,width=0.31\linewidth]{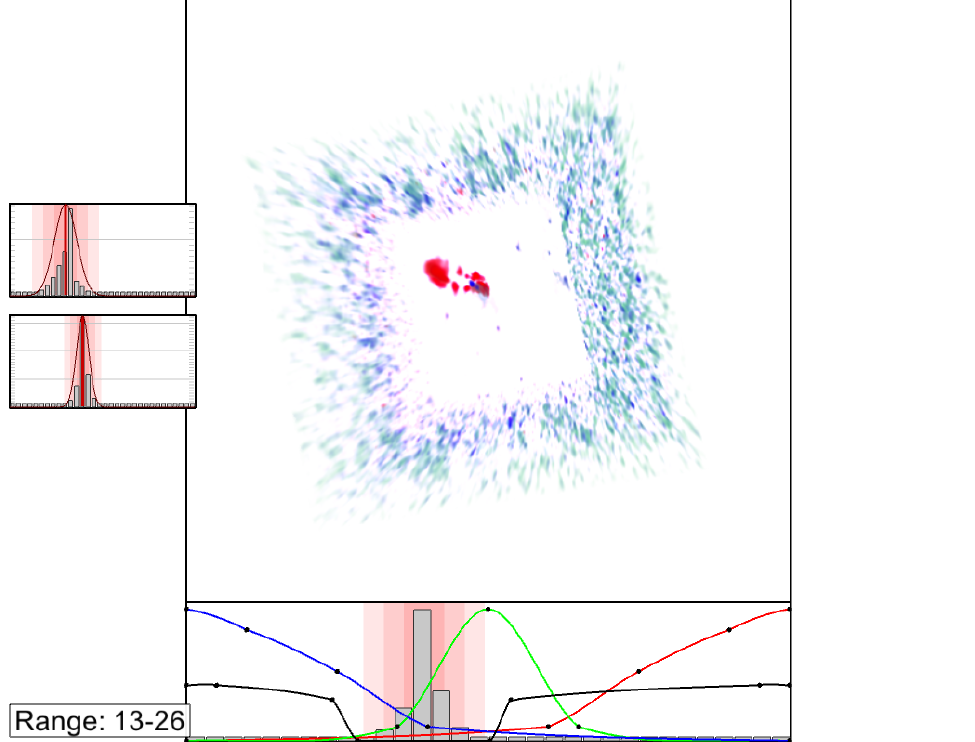}\hfill
    \includegraphics[trim=0 0 75pt 0, clip,width=0.31\linewidth]{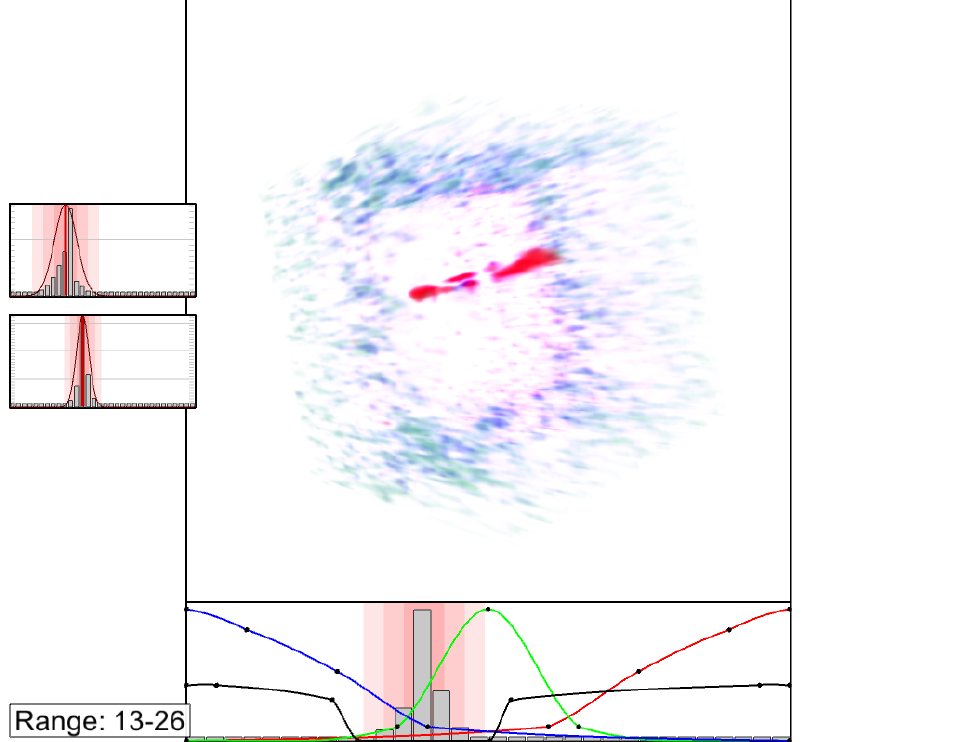}\hfill
    \includegraphics[trim=0 0 75pt 0, clip,width=0.31\linewidth]{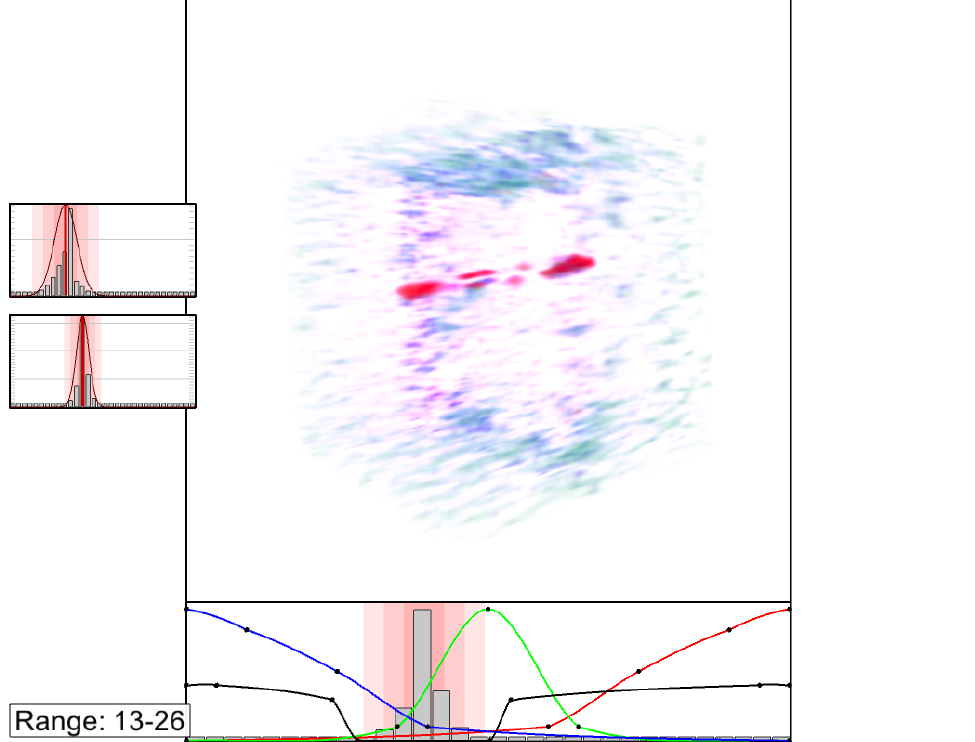}    
    \caption{Result of volume rendering the Ghost of Mirach data set before (top) and after (bottom) using a stack of images with 2D contour trees. The columns show 3 different viewing angles of slices 13-26. The persistent simplification level is $0.00138$. Side views (middle and right) of the rendered volume are blurry due to lower resolution along the spectral dimension compared with the two spatial dimensions.}
    \label{fig:spinningDisk}
\end{figure}

\para{Science Description.} Excited molecular carbon monoxide gas emits light at 230 GHz. The doppler shifts of this line emission can provide information on the motion of molecular gas in the galaxy.  Visualization of the data of NGC404 shows a clear rotating disk located within the central 20 light years of the galaxy (see Fig.~\ref{fig:spinningDisk}).  Similar rotating molecular gas disks have been used to measure the masses of supermassive black holes at the centers of galaxies (e.g.~\cite{BarthDarlingBaker2016, OnishiIguchiDavis2017}). However, the data is noisy, so coherent gas structures are hard to pick out.  NGC 404 presents a special challenge due to the low mass of its black hole~\cite{NguyenSethBrok2017}.  Fortunately, the high angular resolution of ALMA provides the highest sensitivity for measuring the black hole mass.

We can see an example of 4 spectral slices of the data set in Fig.~\ref{fig.anil.original}. In these 4 slices, the bright red spots represents the signal, while most of the remaining patterns represent noise.

\para{Varying Simplification Levels.} Fig.~\ref{fig.anil.varysimp} shows an example of performing simplification on a single 2D spectra (i.e., a single slice along the frequency axis). The noisy structure is captured by the 2D contour tree as many low persistence features (bottom left). Increasing the level of simplification removes much of this noise (right). However, selecting a simplification level that is too aggressive may result in loss of signal (bottom right).

\para{3D Contour Trees.}
Since the spectral data are treated as cubes, our collaborators are interested in the structures that would be found using 3D contour trees. The result of capturing the 3D contour tree, shown in Fig.~\ref{fig.anil.3dvolume}, is both a surprise and a disappointment. Although many critical points are found, the data suffer from \textit{topological pants}---a sphere with three disjoint closed discs removed~\cite{BasmajianSaric2016}.
Essentially, the 3D contours of noisy features form a complex interconnect tubes through the volume that are not physically meaningful. This interferes with the kind of features that a contour tree can identify. The root cause of this is that each of the spectra are processed independently, and thus, there is no correlation between noise patterns across consecutive slices. Simplifying these temporal noise patterns as a whole is not physically meaningful, and they interfere with true features in the data. 

\begin{figure}[!t]
	\centering
	\includegraphics[width=0.48\linewidth]{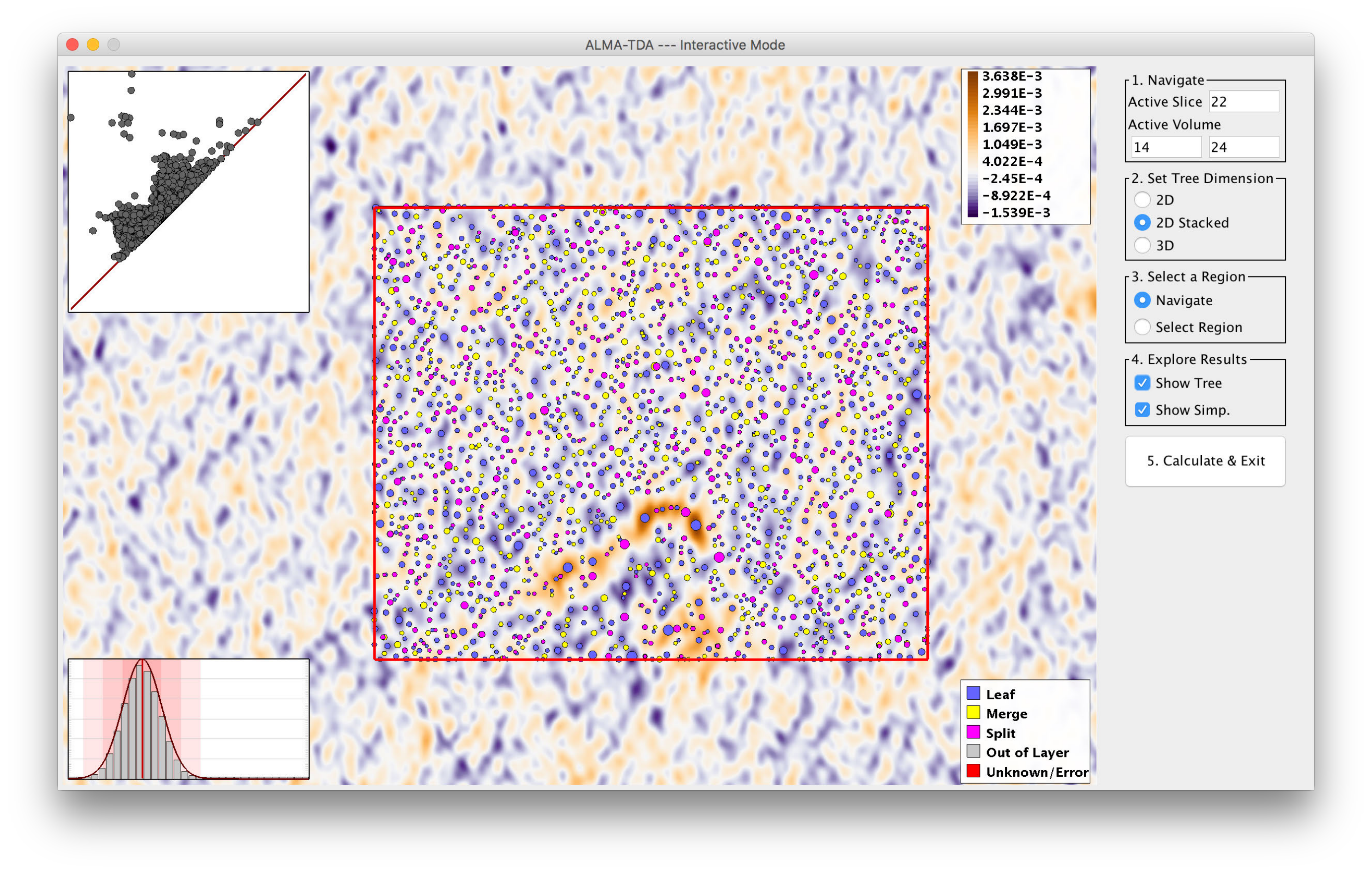}
	\includegraphics[width=0.48\linewidth]{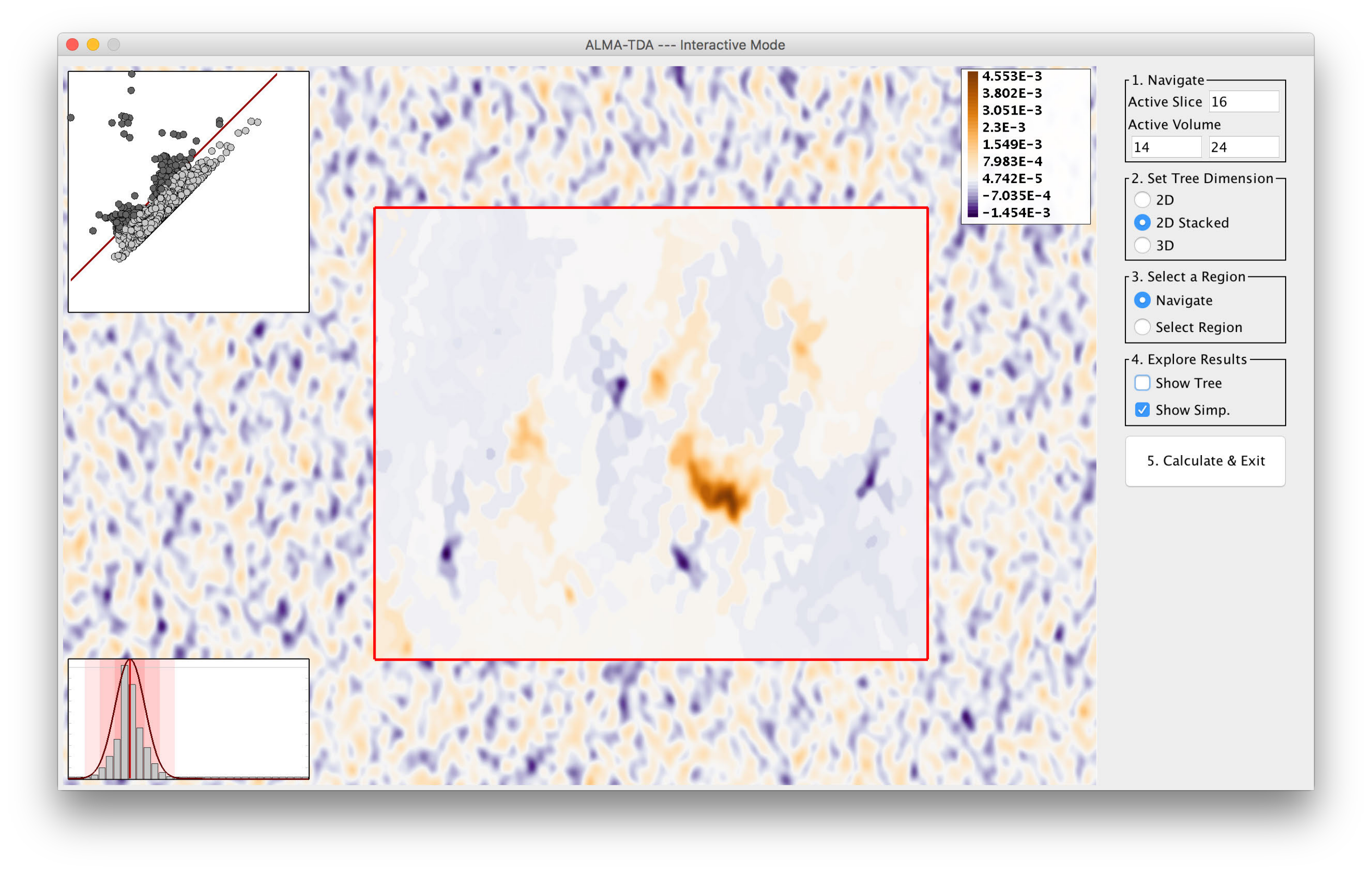}
	\includegraphics[width=0.32\linewidth]{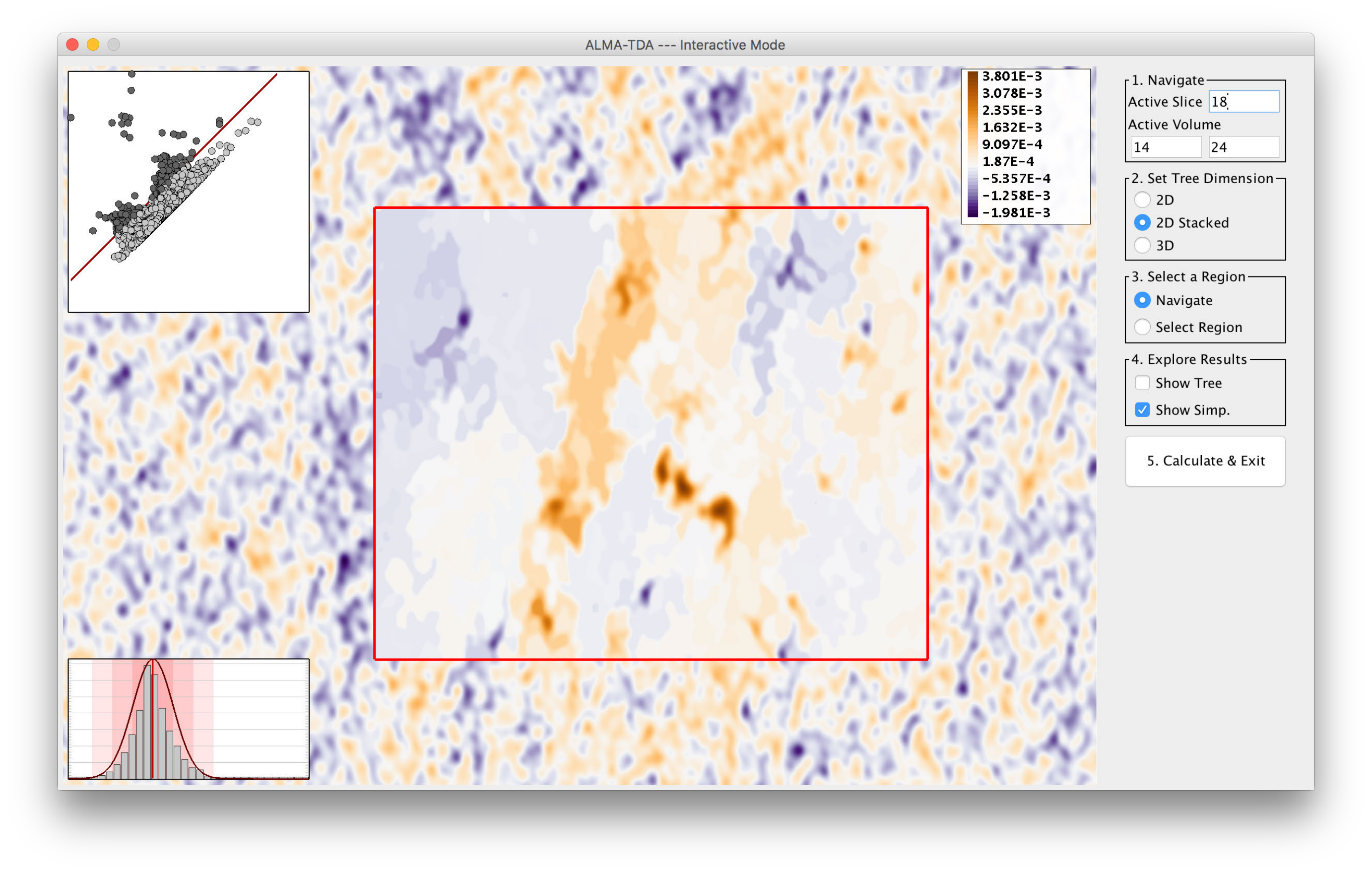}
	\includegraphics[width=0.32\linewidth]{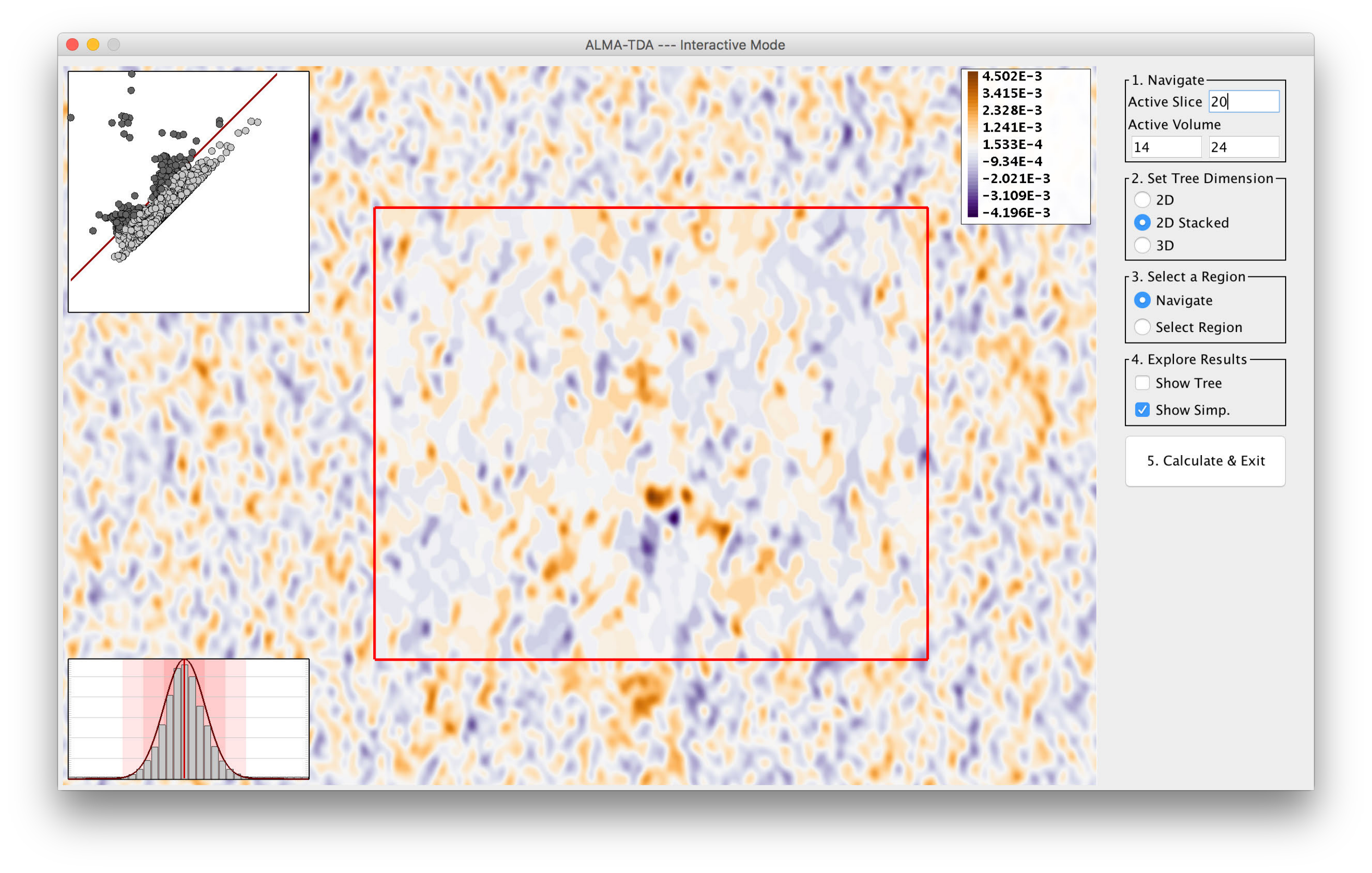}
	\includegraphics[width=0.32\linewidth]{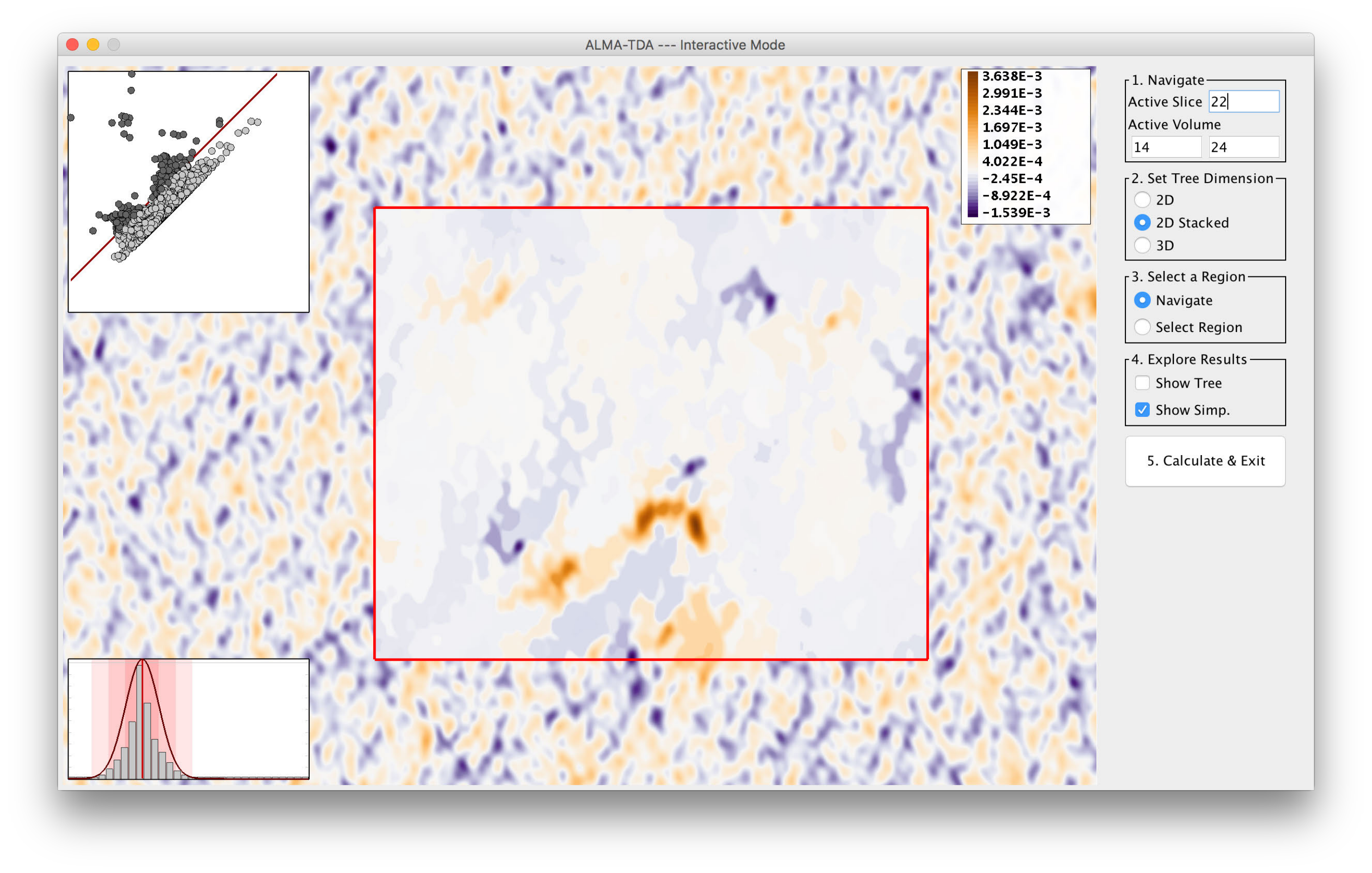}
	\caption{Result of simplifying the Ghost of Mirach data set using a stack of images with 2D contour trees. Top left: Visualization of the 2D contour tree on slice 22. Top right: Simplification of slice 16. Bottom: Simplification of slices 18, 20, \& 22, respectively. The persistent simplification level is $0.00138$. The simplification level is good for all except slice 20 where a more aggressive level of simplification is called for.}
	\label{fig.anil.2dstack}
\end{figure}

\para{2D Contour Tree Stacks.}
On the other hand, the processing of 2D contour trees is highly successful. However, domain scientists still need the ability to process 3D cubes. The obvious solution is to use a series of 2D contour trees to control the simplification. Fig.~\ref{fig.anil.2dstack} shows the result of simplifying a stack of spectra (slices). This example uses a similar level of simplification to the 3D contour tree example in Fig.~\ref{fig.anil.3dvolume}. In our implementation, level of simplification is shared between all slices. This works well for slices 16, 18, and 22 (top right, bottom left, and bottom right, respectively). However, the level of simplification is not aggressive enough for slice 20 (bottom middle). At this point the user could either select a more aggressive simplification, or they could choose to simplify slice 20 separately from the others. Fig.~\ref{fig:spinningDisk} shows the stack of slices 13-26 drawn using a custom-built conventional volume renderer. Despite the natural denoising properties of volume rendering, the results without persistent simplification (Fig.~\ref{fig:spinningDisk} top) are difficult to interpret when compared to those with contour tree stack simplification (Fig.~\ref{fig:spinningDisk} bottom).

\begin{figure}[!t]
	\centering
		\includegraphics[width=0.48\linewidth]{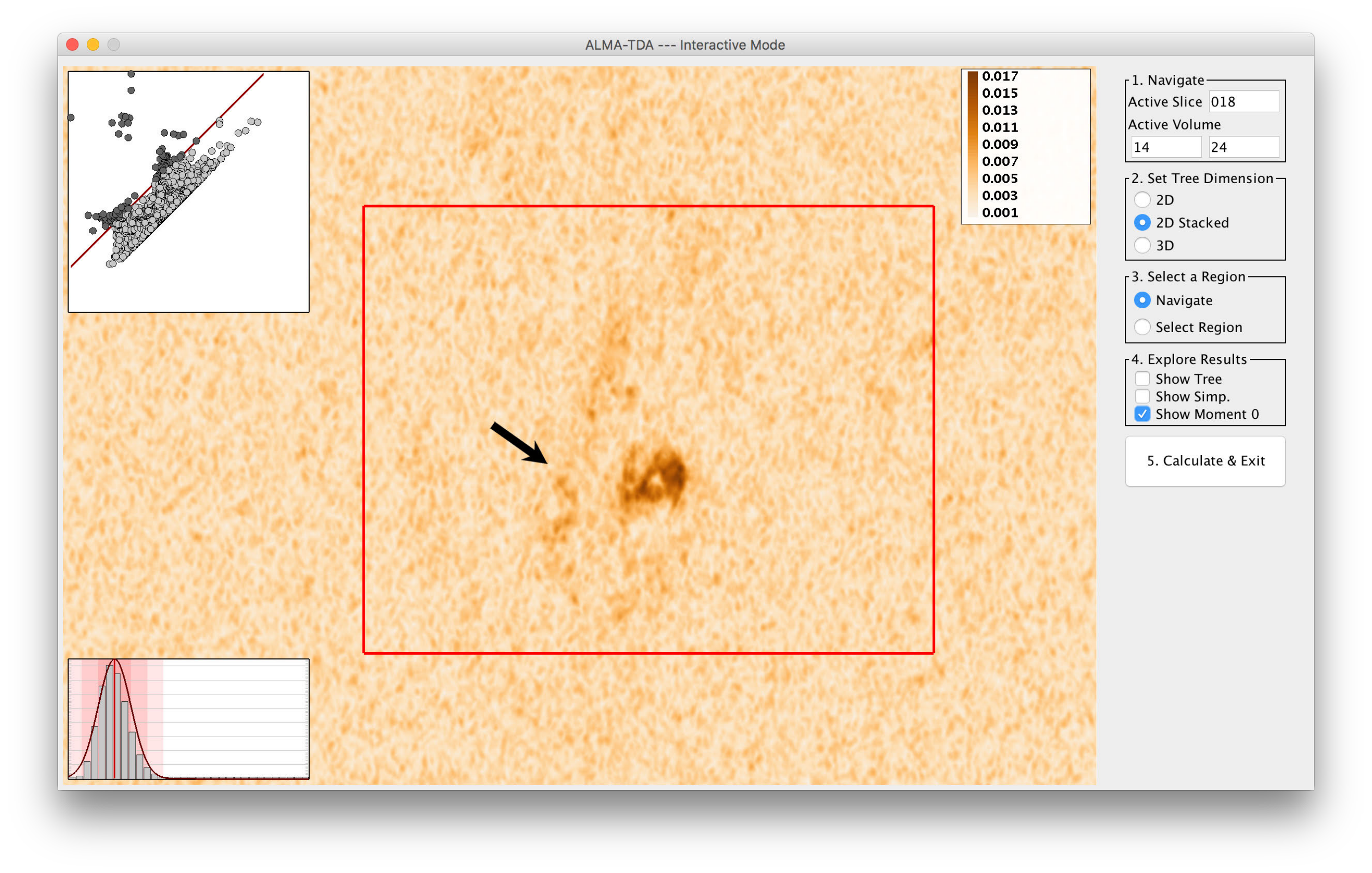}
		\includegraphics[width=0.48\linewidth]{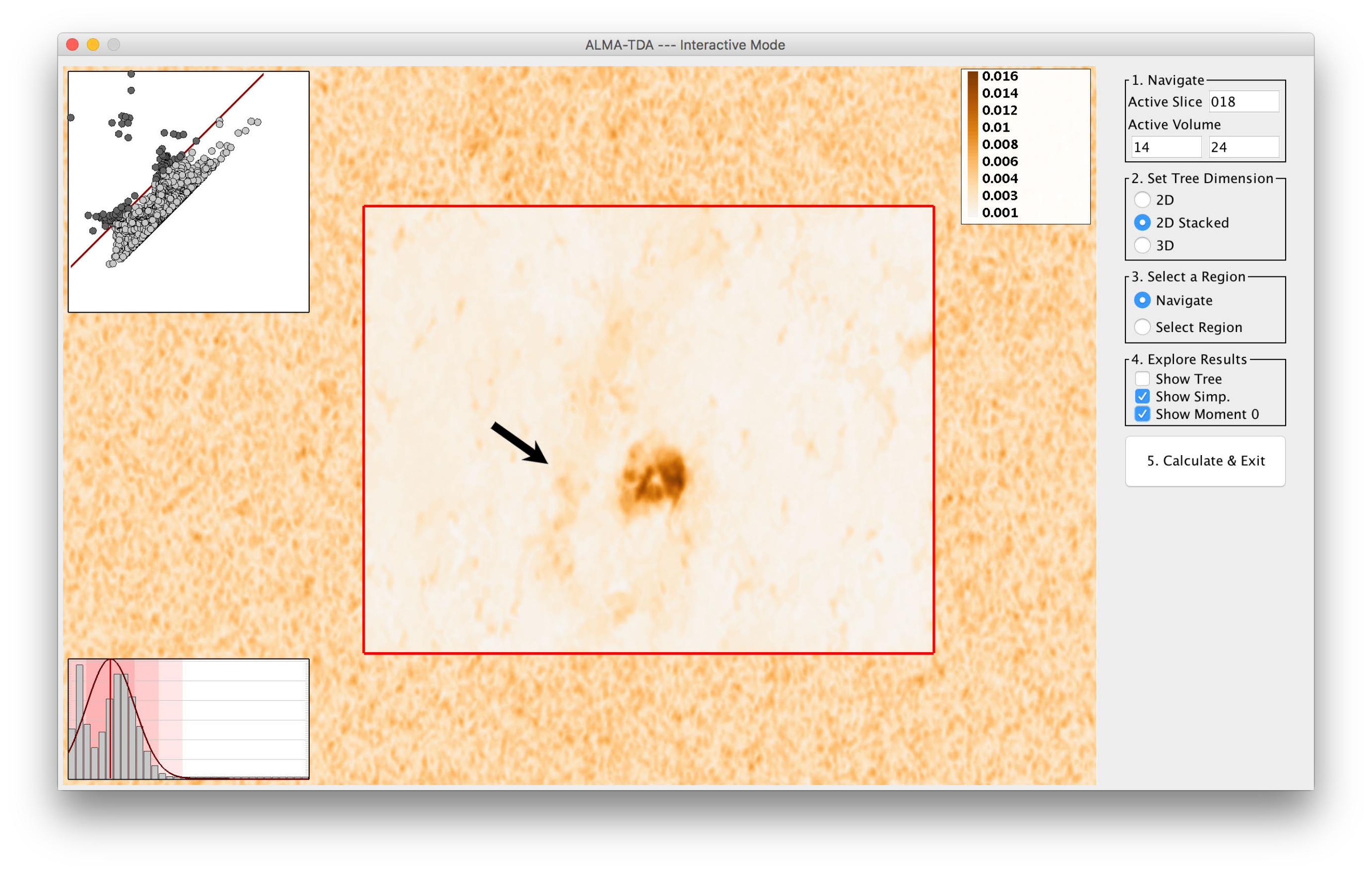}
    
	\caption{Moment 0 analysis of Ghost of Mirach data set between slices 14 and 24 (the range of the signal) using a stack of 2D contour trees. Left: Visualization of moment 0 for original data. Right: Moment 0 results using data with simplification level of $0.0020$.}
	\label{fig.anil.moment0}
\end{figure}

\para{Moment 0 Analysis.}
Astrophysicists often use what is known as \emph{moment analysis} to reduce the 3D spectrum to 2D images. Moment $0$, $1$, and $2$ measure the mass of gas, the direction of gas movement, and the temperature of gas, respectively. They are all integrals across the spectra. To demonstrate the noise reducing power of our approach, we show the result of moment 0 analysis in Fig.~\ref{fig.anil.moment0} on the 2D stack simplification from Fig.~\ref{fig.anil.2dstack}. Moment 0 is calculated as $m_0 = \int I_v$, where $I$ is the intensity for a given spectra $v$. By removing the noise from each of the layers, the resulting moment map is significantly less noisy making the signal itself very apparent. Our collaborator also finds the dim feature pointed to by the arrow very interesting. 
He and his collaborators have been actively debating whether this structure is signal or a data processing artifact. 
Nevertheless, our approach retains it as a signal, and we are excited to see how our results generate further conversations regarding the data.

\subsection{CMZ Data Set}

The CMZ data are a $^{13}$CO 2-1 image of the Central Molecular Zone (CMZ) of the galaxy (data are published in~\cite{ginsburg2016dense}).
The data cube is approximate 500MB with resolution of 1150x200 in the spatial domain and 500 in the spectral domain (i.e., 500 slices). We look at 100 slices of a region with resolution about 300x200.

\begin{figure}[!t]
	\centering
		\fbox{\includegraphics[trim = 305pt 225pt 330pt 230pt, clip,width=0.1725\linewidth]{\imgExt{adam_3_45/org_100}}} \hfill
        \fbox{\includegraphics[trim = 305pt 225pt 330pt 230pt, clip,width=0.1725\linewidth]{\imgExt{adam_3_45/org_120}}}
        \hfill
		\fbox{\includegraphics[trim = 305pt 225pt 330pt 230pt, clip,width=0.1725\linewidth]{\imgExt{adam_3_45/org_140}}} \hfill
		\fbox{\includegraphics[trim = 305pt 225pt 330pt 230pt, clip,width=0.1725\linewidth]{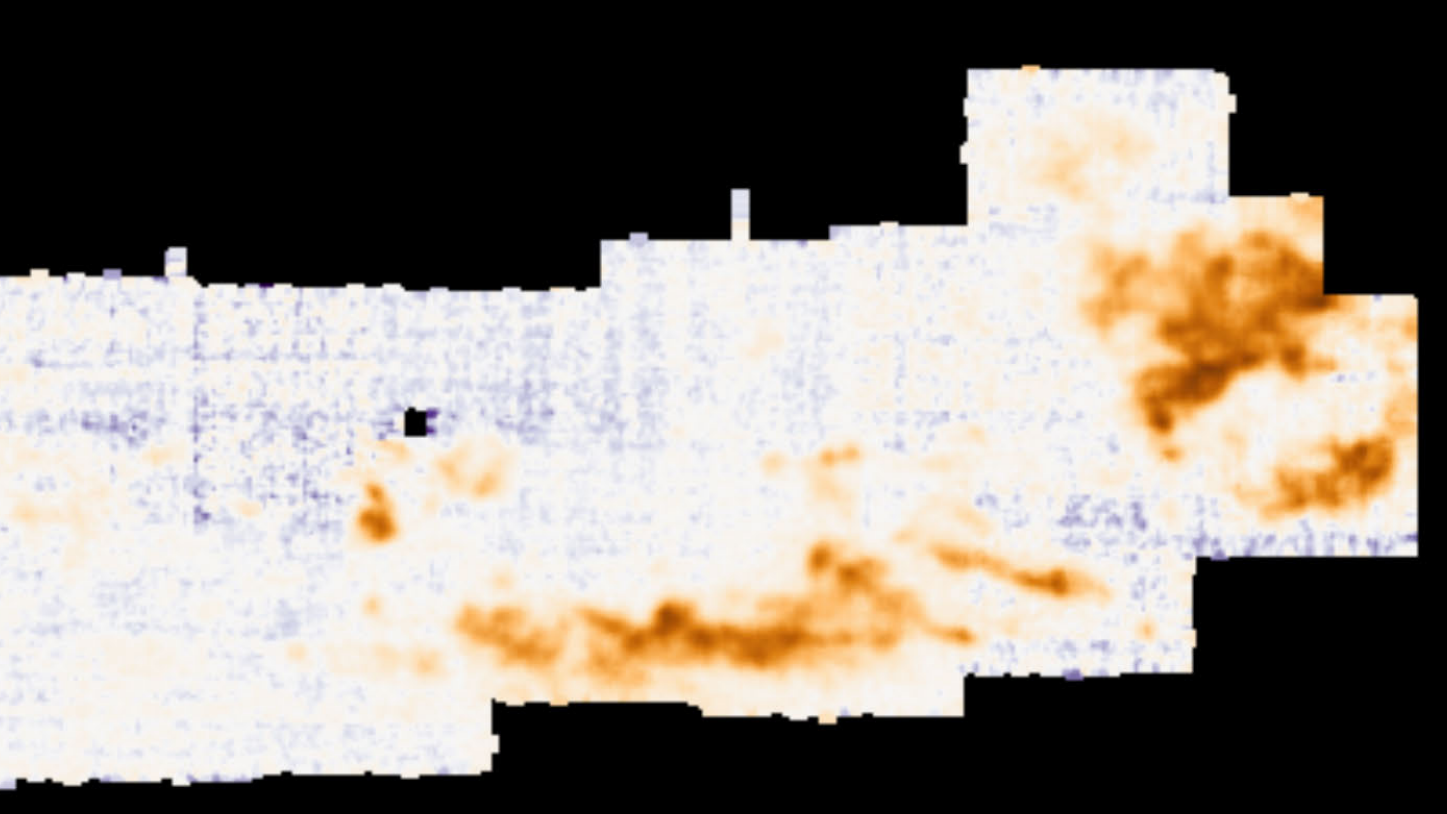}} \hfill
		\fbox{\includegraphics[trim = 305pt 225pt 330pt 230pt, clip,width=0.1725\linewidth]{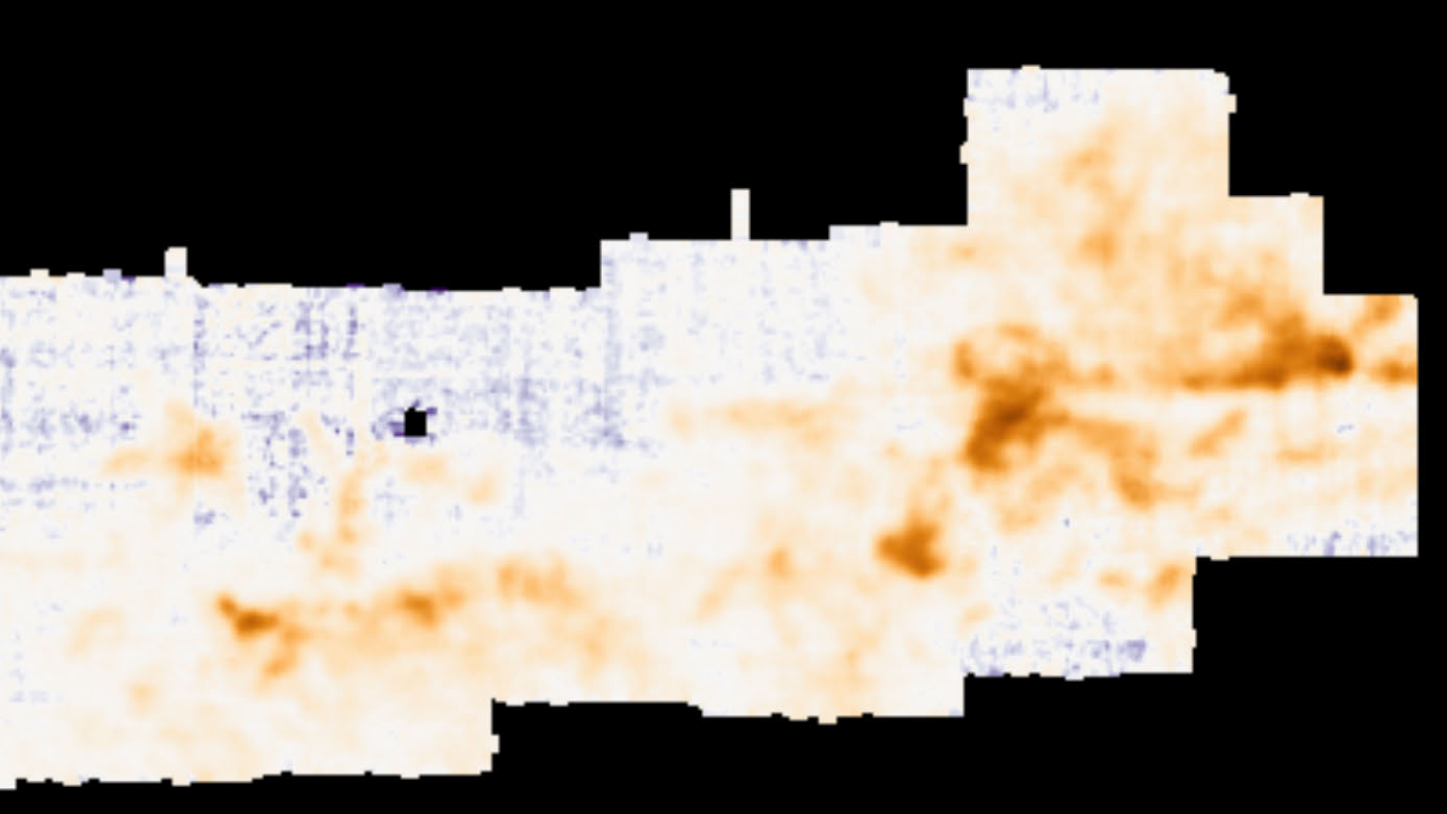}}
        
		\vspace{5pt}
		\fbox{\includegraphics[trim = 305pt 225pt 330pt 230pt, clip,width=0.1725\linewidth]{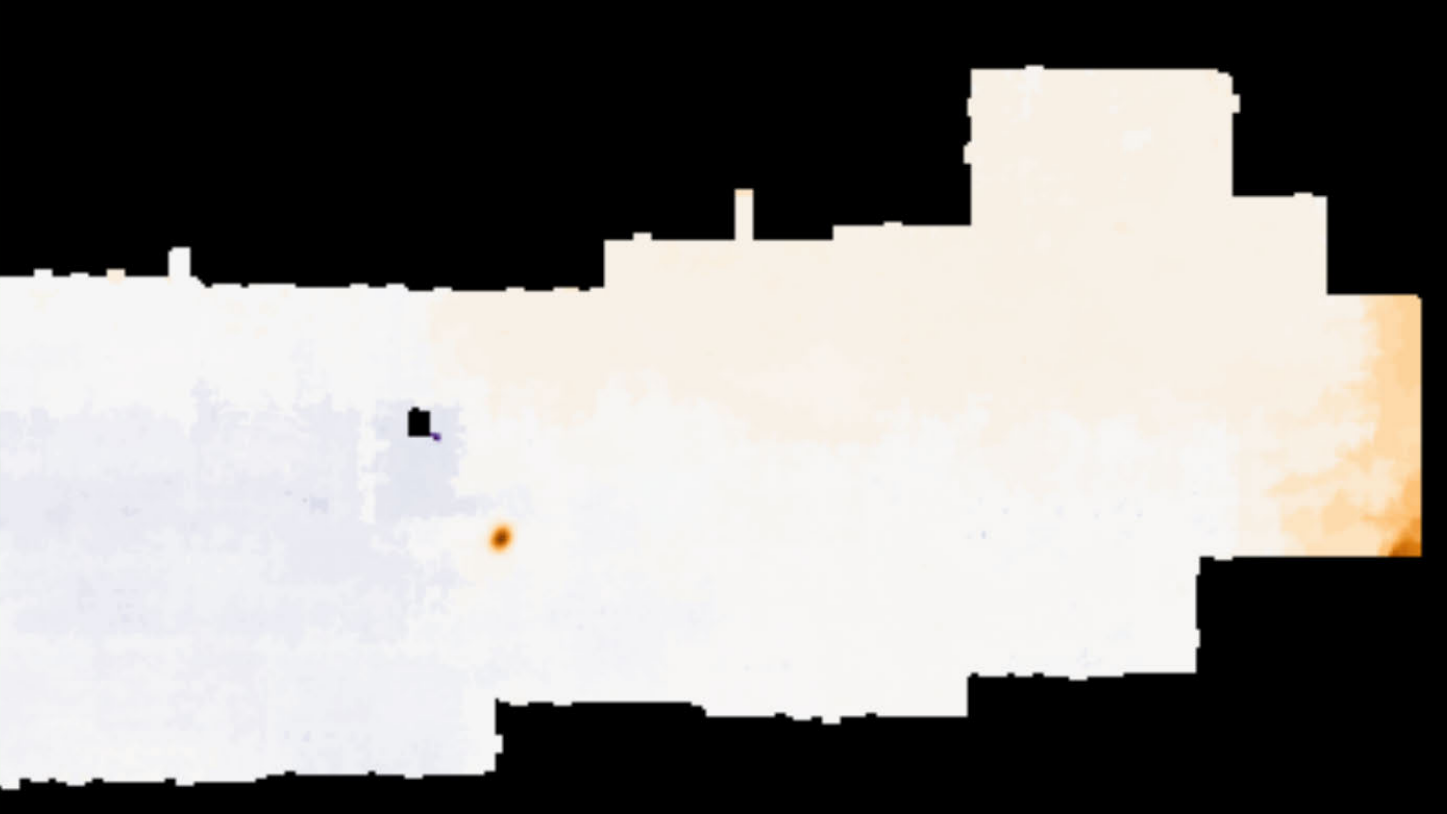}} \hfill
		\fbox{\includegraphics[trim = 305pt 225pt 330pt 230pt, clip,width=0.1725\linewidth]{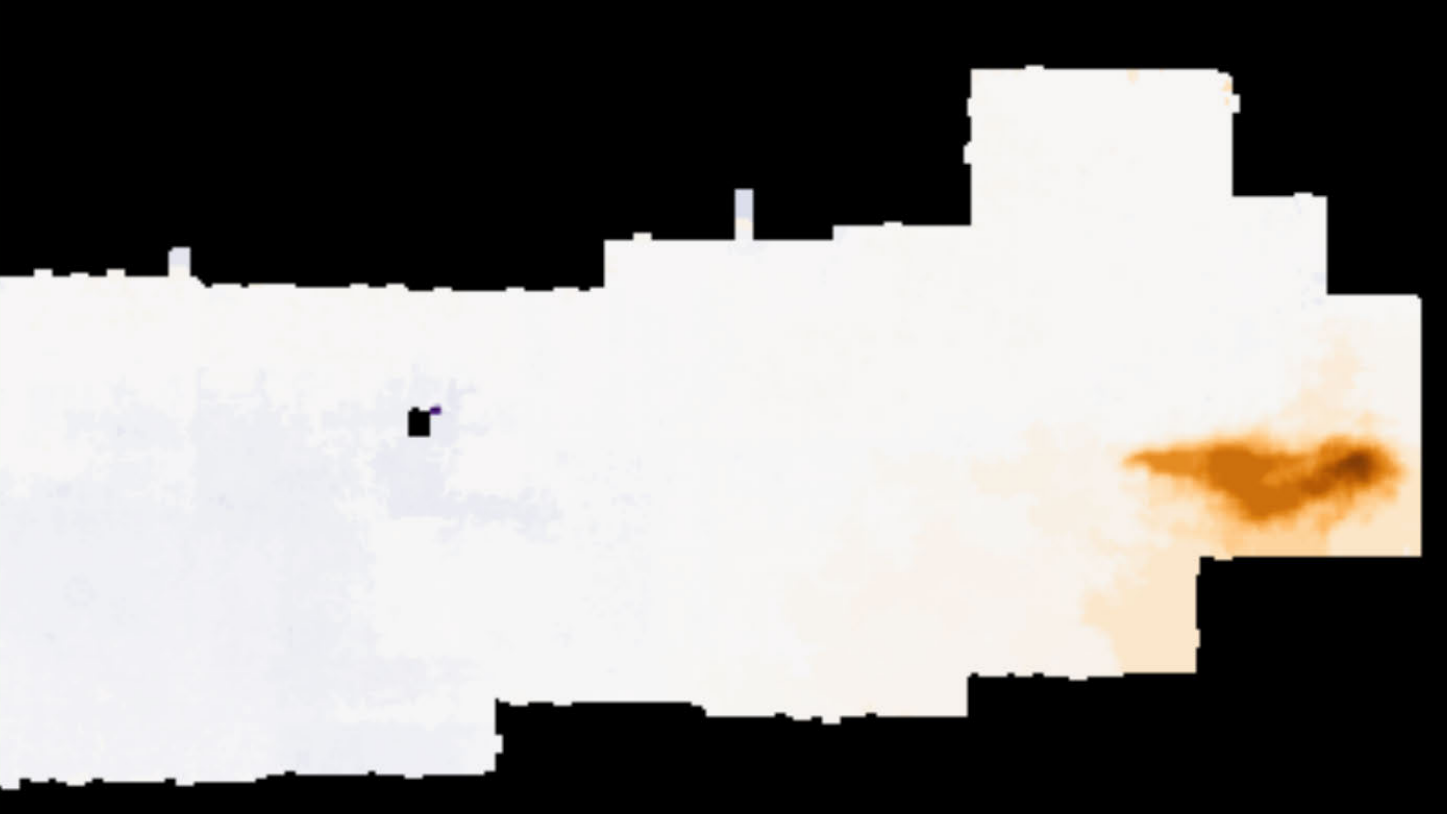}} \hfill
		\fbox{\includegraphics[trim = 305pt 225pt 330pt 230pt, clip,width=0.1725\linewidth]{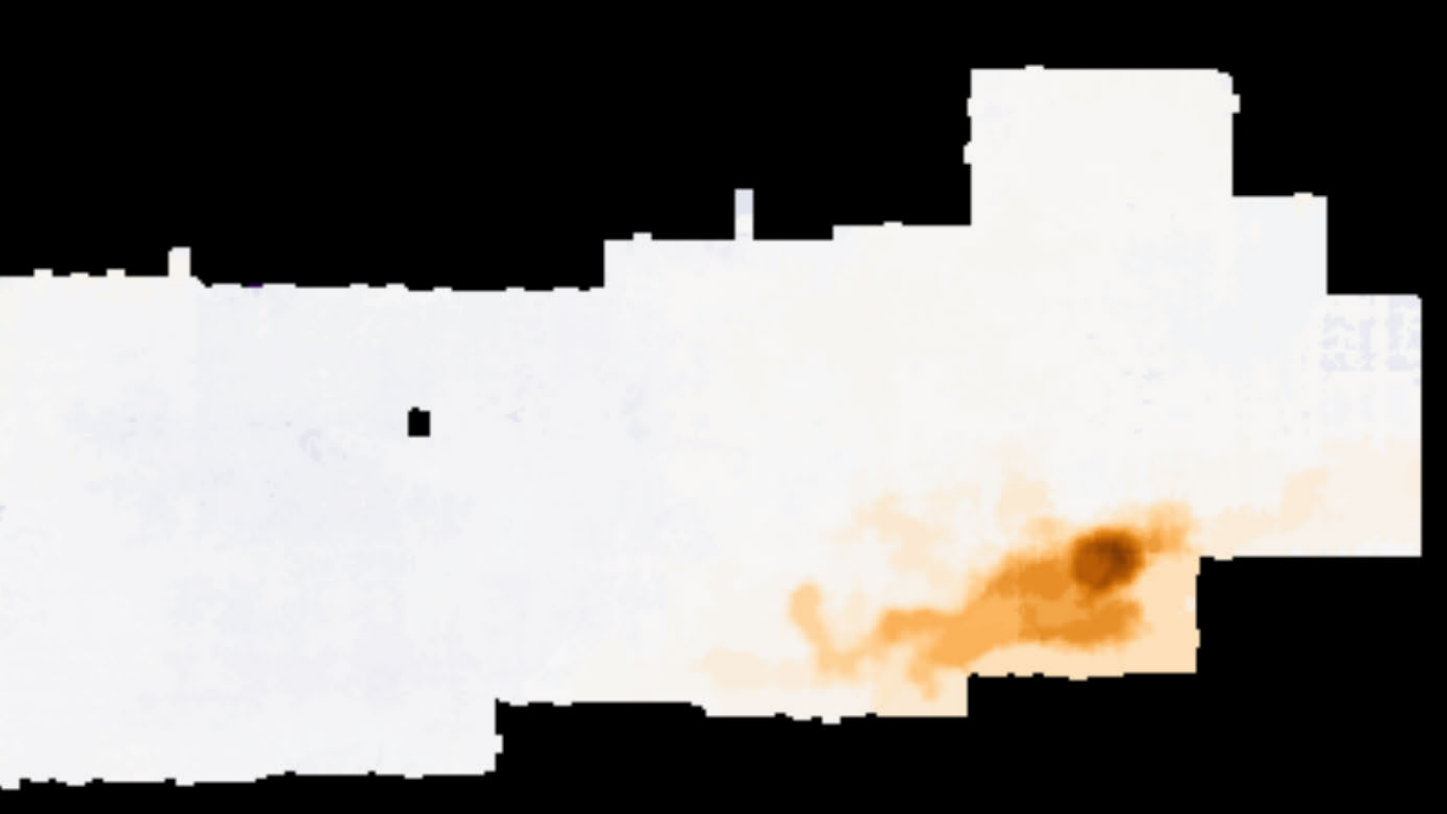}} \hfill
		\fbox{\includegraphics[trim = 305pt 225pt 330pt 230pt, clip,width=0.1725\linewidth]{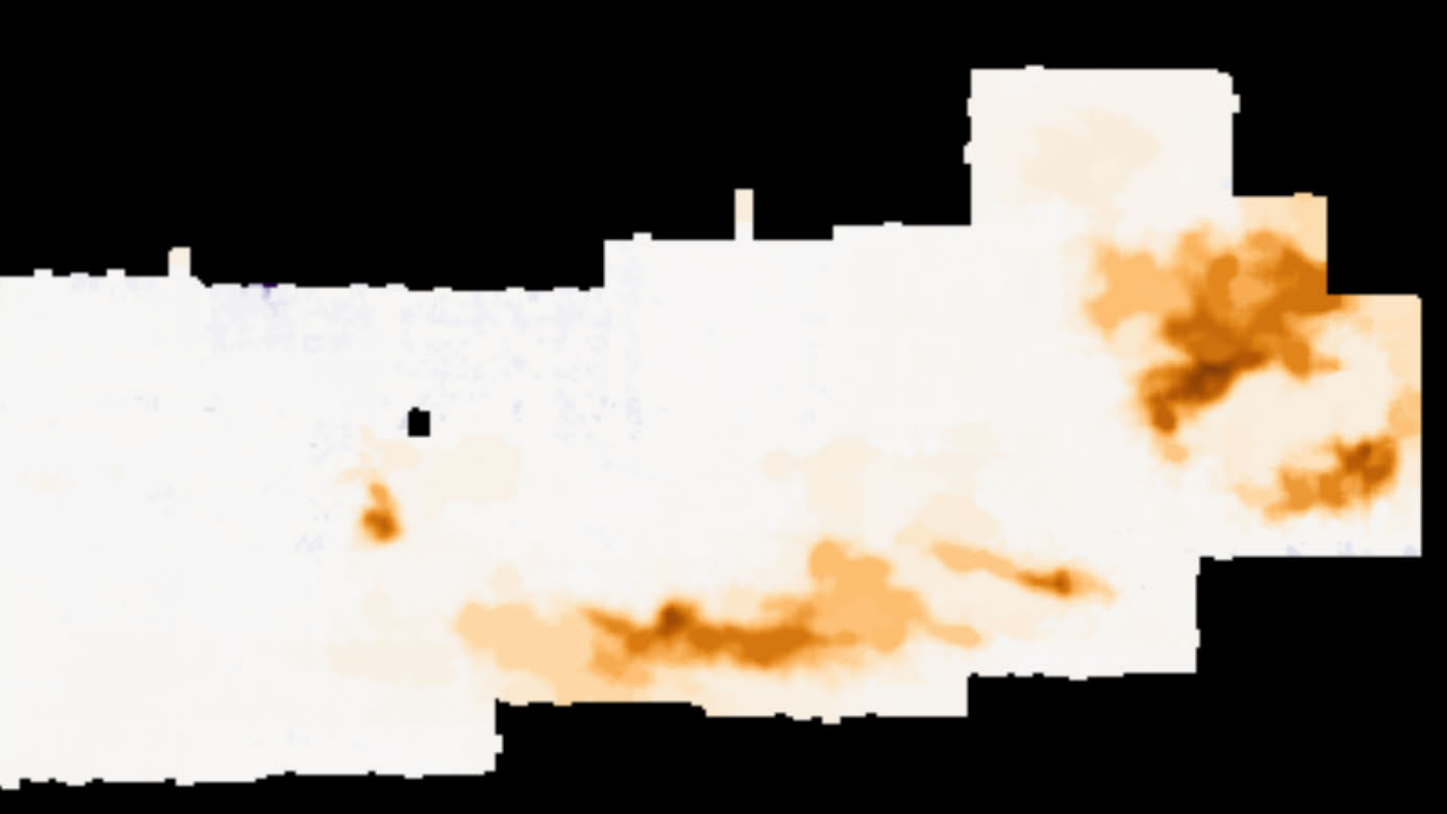}} \hfill
        \fbox{\includegraphics[trim = 305pt 225pt 330pt 230pt, clip,width=0.1725\linewidth]{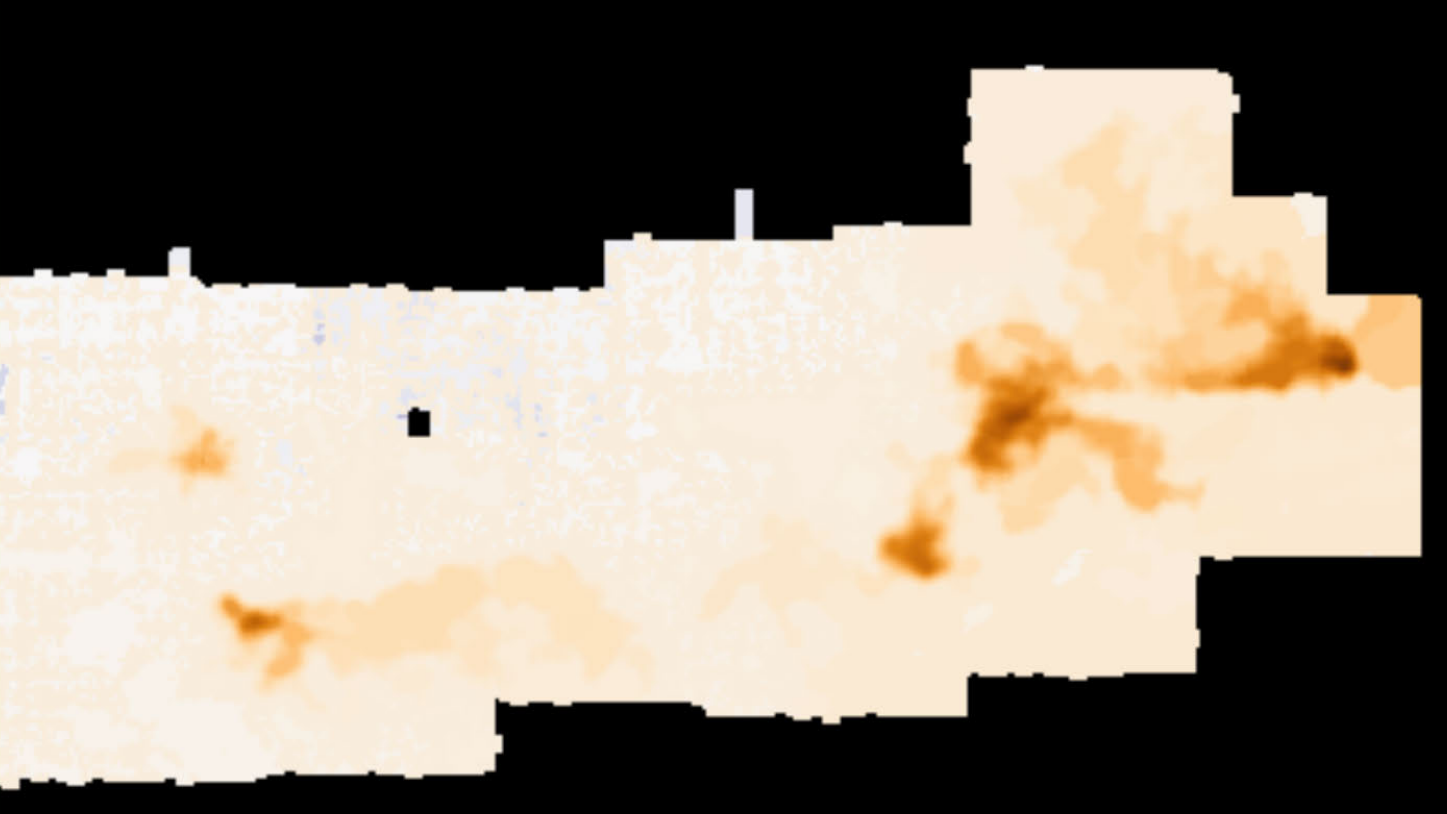}}

	\caption{Visualizations of selected slices from the range $100$ to $200$ of the CMZ data.  Top: Slices $100$, $120$, $140$, $160$, and $180$ before simplification, respectively. Bottom: Slices $100$, $120$, $140$, $160$, and $180$ after simplification, respectively. The simplification level used is $3.45$.}
	\label{fig.adam.slices}
\end{figure}

\para{Science Description.}
The cube shows the low-density molecular gas in the Galaxy's center,
with higher intensities generally indicating that there is more gas
moving at a particular velocity along each line of sight.  It contains
highly turbulent gas with properties that are very different than the
rest of the Galaxy.  Domain scientists use these data to measure the structure of the
interstellar medium, which is important for determining how stars are
formed and how galaxies evolve.  Because the gas they are seeing is in
diffuse clouds that do not have well-defined edges, signal
identification is a critical component in improving their understanding
of how the gas changes states.  Identifying structures in the gas is
useful for determining how turbulent it is on different scales, which
plays a key role in may star formation theories.

\begin{figure}[!b]
	\centering
 		\fbox{\includegraphics[width=0.475\linewidth]{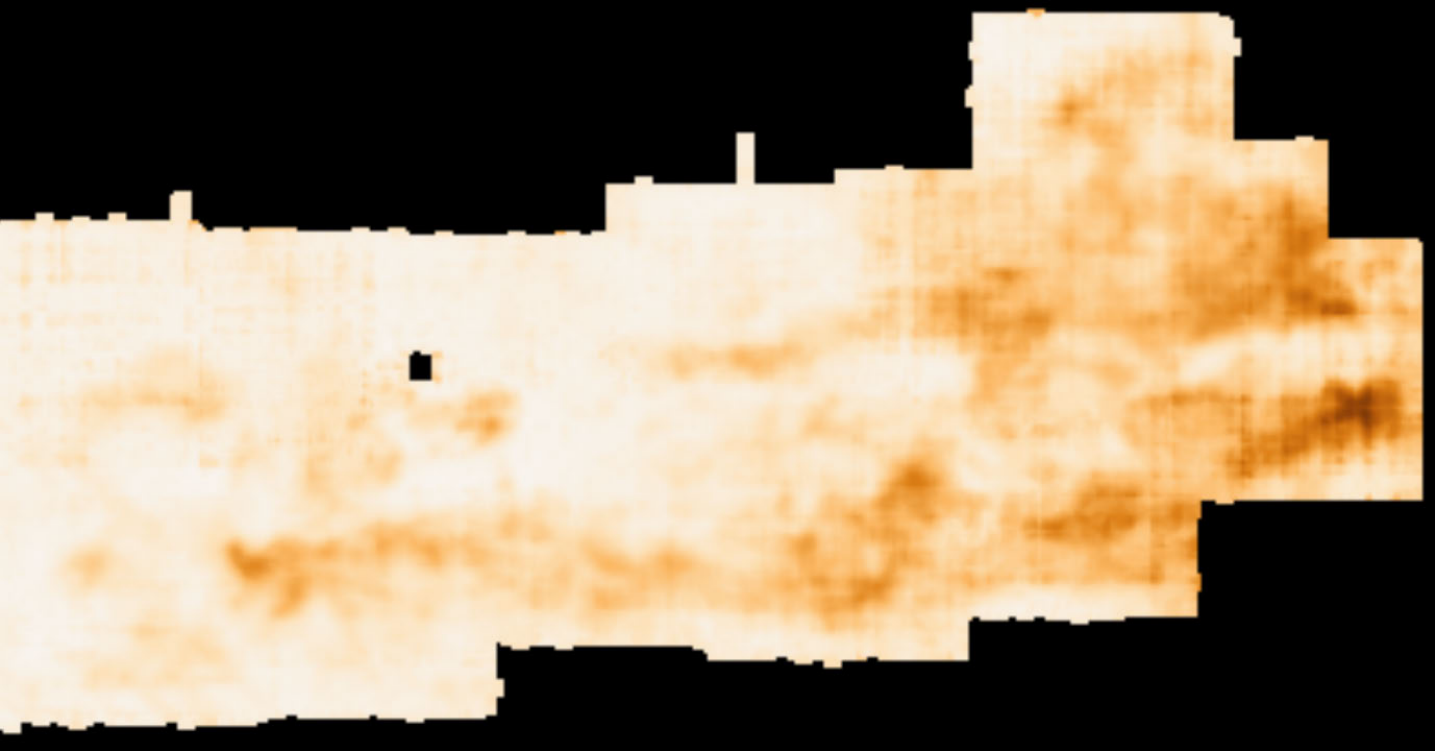}}
       	\fbox{\includegraphics[width=0.475\linewidth]{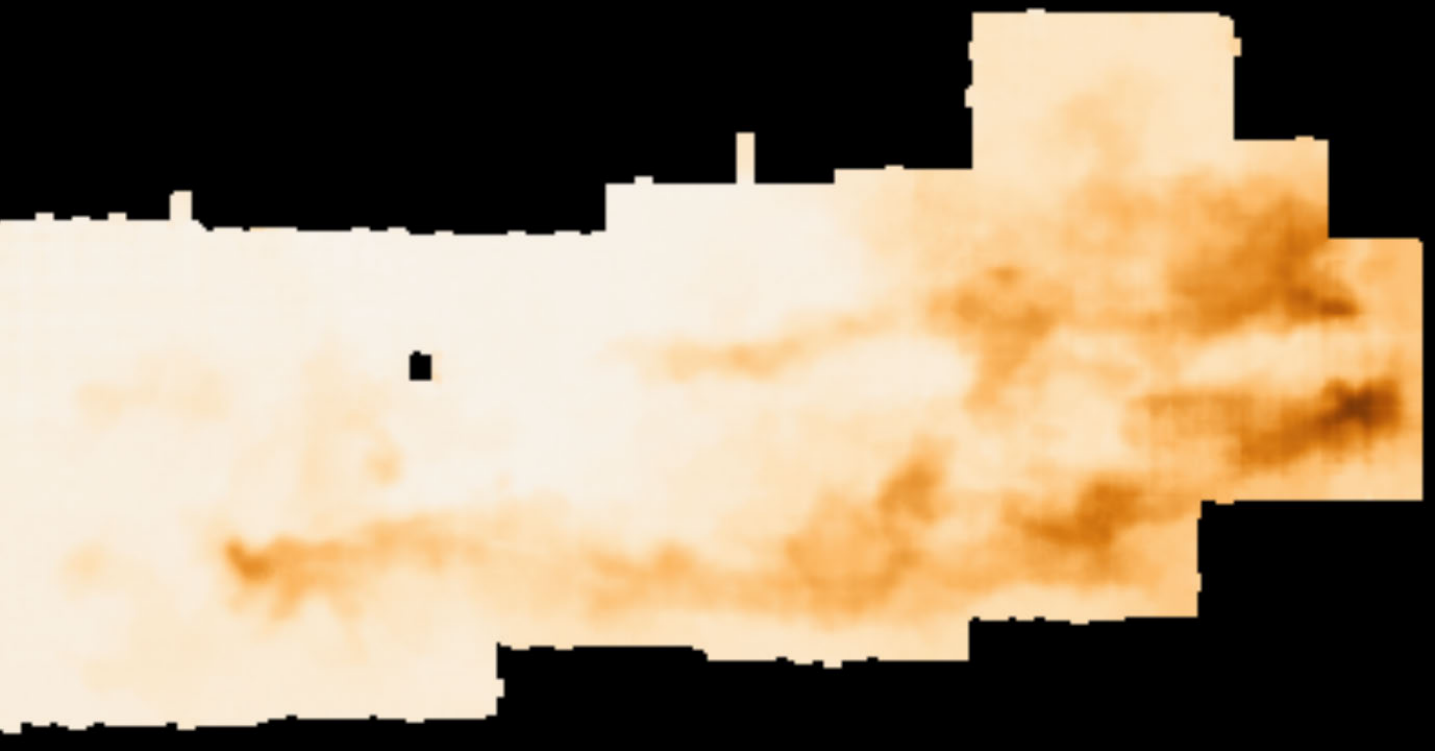}}
   		\fbox{\includegraphics[width=0.475\linewidth]{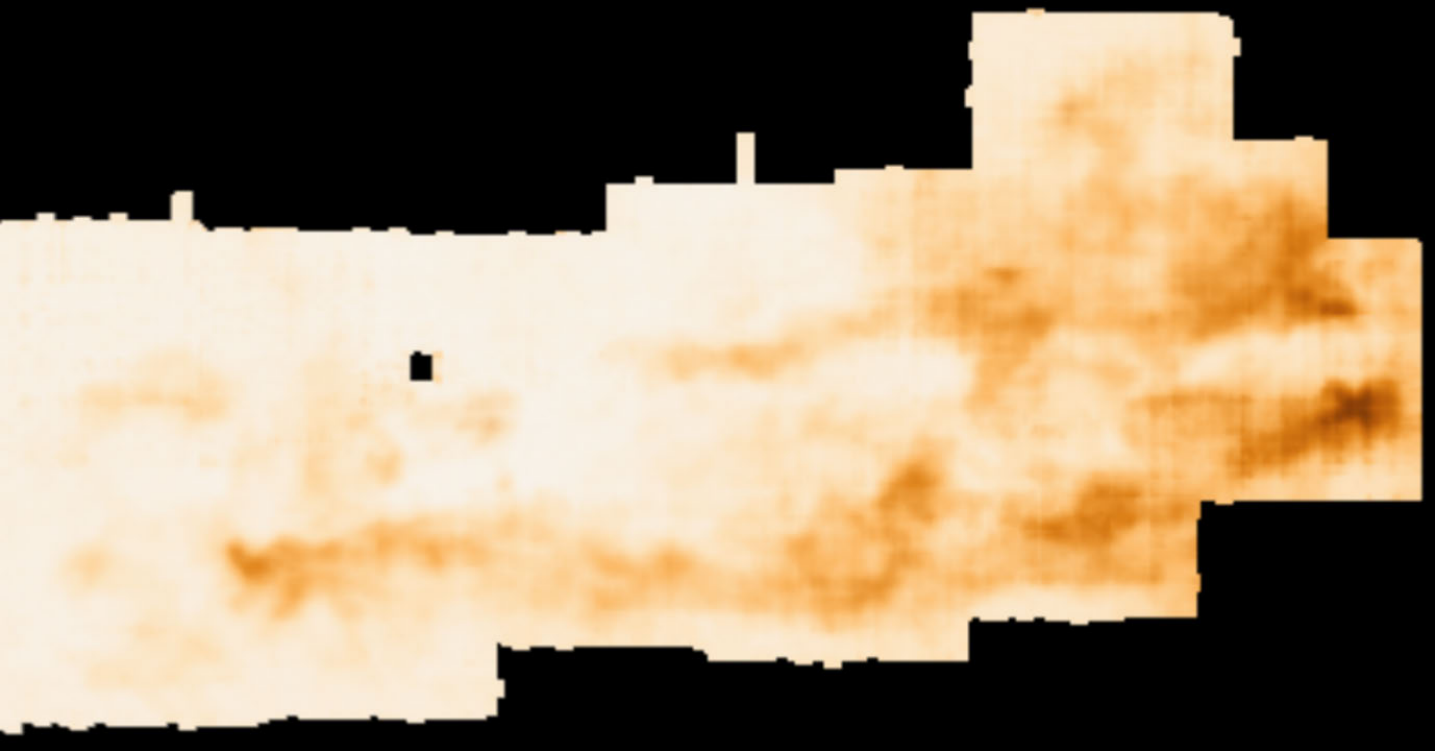}}
   		\fbox{\includegraphics[width=0.475\linewidth]{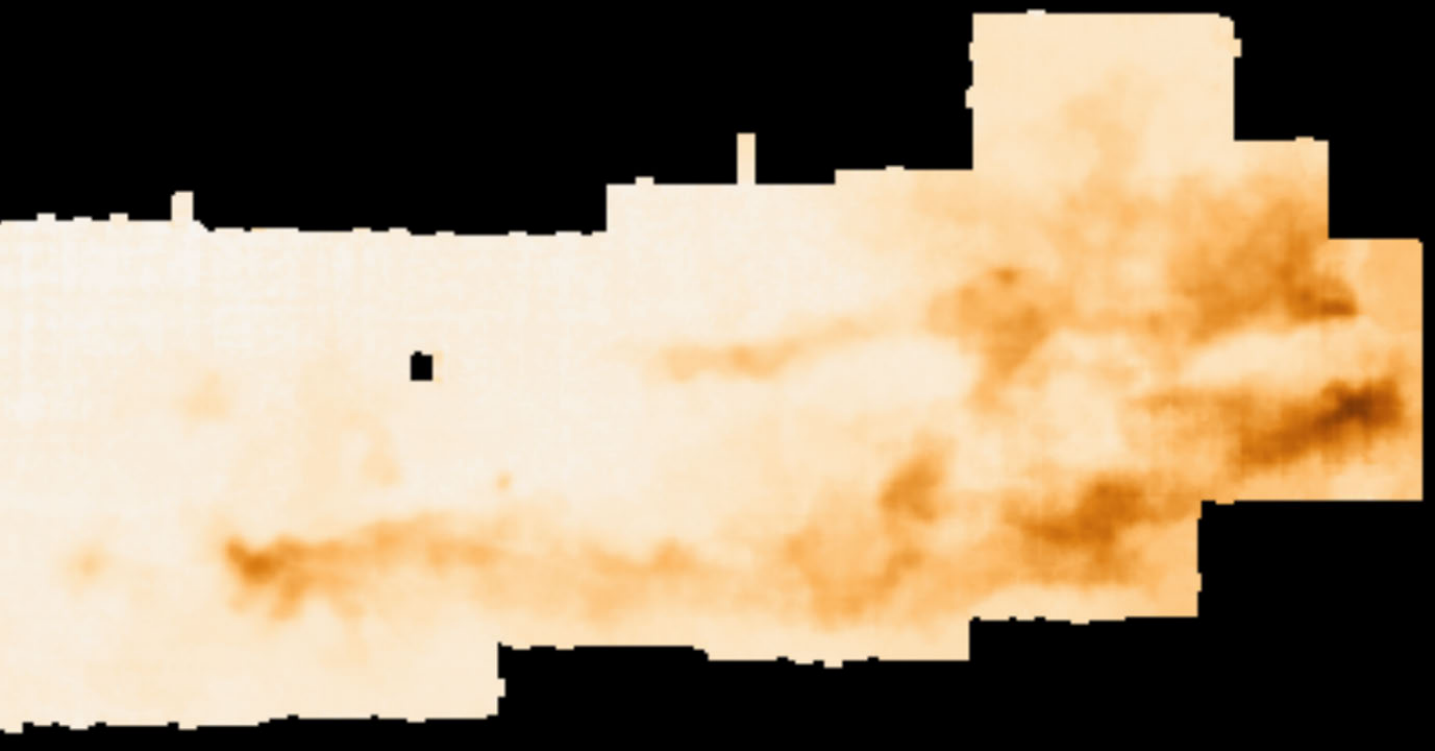}}
	\vspace{-4mm}
	\caption{Visualizations of moment $0$ for slices $100$ to $200$ of the CMZ data set computed using 2D contour trees. Top left: Moment $0$ on the original data. Top right: Moment 0 on all 100 simplified slices. Bottom left: Moment 0 only using every 4th slice. Bottom right: Moment $0$ only using every 8th slice. The simplification level used is $3.45$.}    
	\label{fig.adam.moment}
\end{figure} 



\para{Denoising Slices.}
Fig.~\ref{fig.adam.slices} shows a number of slices denoised. The signal to noise ratio in this data set is much better than the previous ones. Nevertheless, many low persistent features have been removed using our approach.

\para{Denoising for Moment Analysis.}
Deep cubes (those with many slices) such as this one are often created in order to mitigate the impact of noise during moment analysis -- by more densely sampling the frequency domain, noise from any single slice has a smaller impact in the output. However, creating deep cubes such as this is computationally and manpower expensive. NRAO has significant human and computational infrastructure dedicated to generating data cubes from the raw data captured by radio telescopes. By providing strong denoising capabilities, data cubes can be sampled at lower spectral frequencies and still produce similar moment maps. See Fig.~\ref{fig.adam.moment} for an example. Here, the top shows the moment map on the original data. Then, moment maps are shown that are calculated using every slice ($100$ total), every 4th slices ($25$ total), and every 8th slice ($12$ total). The results using fewer slices are virtually indistinguishable from the version using all $100$ slices.

%% file: sec-ALMA-discussion.tex
\section{Discussion}
\label{sec:discussion}

In this feasibility study, we focus on persistence-based simplification of ALMA data cubes.
Our application development process focuses on the usability objectives of collaborators, \emph{simplicity}, \emph{integrability}, and \emph{reproducibility}, and we recommend these design objectives for anyone else wishing to collaborate with astrophysicists.

Despite our initial inclination to build a large scale visualization system, we find  that this is unnecessary given the existing array of visualization options. 
Instead, what is needed is a simple and compact tool to understand the impact of parameter selection on the data via visualization. 
Parameter selection is not intuitive to new users. Without the visualization of the parameter selection, that intuition is relatively difficult to build. Nevertheless, once the selection is complete, the visualization and data processing can be easily reproduced using the information retained via the command line interface.


Thus far, reception of our approach has been good. Virtually everyone who has seen the results are impressed, for some, almost to the point of skepticism. Public outreach with such a new tool using unfamiliar techniques remains challenging. Among astrophysicists, there is a desire to understand both the tool and the underlying technique, and given the complexities of topological data analysis, this can be a challenging, but potentially transformative undertaking.

